# MULTILEVEL EVOLUTIONARY DEVELOPMENTAL OPTIMIZATION (MEDO): A THEORETICAL FRAMEWORK FOR UNDERSTANDING PREFERENCES AND SELECTION DYNAMICS

Adam Safron
Indiana University

## Abstract

What is motivation and how does it work? Where do goals come from and how do they vary within and between species and individuals? Why do we prefer some things over others? MEDO is a theoretical framework for understanding these questions in abstract terms, as well as for generating and evaluating specific hypotheses that seek to explain goal-oriented behavior. MEDO views preferences as selective pressures influencing the likelihood of particular outcomes. With respect to biological organisms, these patterns must compete and cooperate in shaping system evolution. To the extent that shaping processes are themselves altered by experience, this enables feedback relationships where histories of reward and punishment can impact future motivation. In this way, various biases can undergo either amplification or attenuation, resulting in preferences and behavioral orientations of varying degrees of inter-temporal and inter-situational stability. MEDO specifically models all shaping dynamics in terms of natural selection operating on multiple levels—genetic, neural, and cultural—and even considers aspects of development to themselves be evolutionary processes. Thus, MEDO reflects a kind of generalized Darwinism, in that it assumes that natural selection provides a common principle for understanding the emergence of complexity within all dynamical systems in which replication, variation, and selection occur. However, MEDO combines this evolutionary perspective with economic decision theory, which describes both the preferences underlying individual choices, as well as the preferences underlying choices made by engineers in designing optimized systems. In this way, MEDO uses economic decision theory to describe goal-oriented behaviors as well as the interacting evolutionary optimization processes from which they emerge. *(Please note: this manuscript was written and finalized in 2012.)*

## TABLE OF CONTENTS

## INTRODUCTION

*"The aim of scientific explanation throughout the ages has been unification, i.e. the comprehending of a maximum of facts and regularities in terms of a minimum of theoretical concepts and assumptions."*
        –Herbert Feigl (1902-1988)

What is motivation and how does it work? Where do goals come from and how do they vary within and between species and individuals? Why do we prefer some things and not others? *Multilevel Evolutionary Developmental Optimization* (*MEDO*) is a theoretical framework for understanding these questions in abstract terms, as well as for generating and evaluating specific hypotheses that seek to explain the characteristics, origins, and transformations of goal-oriented behavior.

MEDO views preferences as selective processes influencing the likelihood of particular outcomes, which are more or less consistent with the dynamics underlying those influences. With respect to biological organisms, since different preferences direct individuals towards goals that can be compatible or incompatible to varying degrees, these patterns must compete and cooperate in shaping system evolution. To the extent that shaping processes are themselves altered by experience, this enables feedback relationships where histories of reward and punishment can impact future motivation. In this way, various biases can undergo either amplification or attenuation of the degree to which different outcomes are valued, potentially resulting in enduring preferences and orientations.

MEDO specifically models all shaping dynamics in terms of natural selection operating on multiple levels—genetic, neural, and cultural—and even considers aspects of development to themselves be evolutionary processes. Thus, MEDO reflects a kind of *generalized Darwinism*, in that it assumes that natural selection provides a common principle for understanding the emergence of complexity within all dynamical systems in which replication, variation, and selection occur, regardless of the specific details of mechanistic implementation (Aldrich et al., 2008; Campbell, 2009; R. Dawkins, 1983; Nelson, 2007).

However, MEDO combines this evolutionary perspective with *economic decision theory*, which describes both the preferences underlying individual choices, as well as the preferences underlying choices made by engineers in designing optimized systems (Berger, 1985; Coello, Dehuri, & Ghosh, 2009; Davies, Watson, Mills, Buckley, & Noble, 2011; Davies et al., 2011; Keeney & Raiffa, 1993; Lewis, Chen, & Schmidt, 2006; Marshall, 1925; Pine et al., 2009; Von Neumann & Morgenstern, 1944). In this way, MEDO uses economic decision theory to describe goal-oriented behaviors, as well as the interacting evolutionary optimization processes from which they emerge.

## MULTILEVEL EVOLUTIONARY DEVELOPMENTAL OPTIMIZATION (MEDO): A THEORETICAL FRAMEWORK COMBINING ECONOMIC DECISION THEORY AND GENERALIZED DARWINISM

### MEDO: PREFERENCES AS SELECTIVE DYNAMICS; SELECTIVE DYNAMICS AS PREFERENCES

#### ECONOMIC DECISION THEORY

##### ECONOMIC DECISION THEORY, EXPECTED UTILITY, AND PREFERENCES

*"Utility is taken to be correlative to Desire or Want. It has been already argued that desires cannot be measured directly, but only indirectly, by the outward phenomena to which they give rise: and that in those cases with which economics is chiefly concerned the measure is found in the price which a person is willing to pay for the fulfillment or satisfaction of his desire."*

–Alfred Marshall (1842-1924)

In order to understand goal-oriented behavior, MEDO utilizes and expands upon conceptual frameworks previously developed for analyzing economic choices (Marshall, 1925; Von Neumann & Morgenstern, 1944). *Economic decision theory* (*EDT*) is the dominant theoretical paradigm of microeconomics, but it has also been extended to political science, sociology, psychology, neuroscience, and even philosophy (Scott, 2010). At its core, EDT explains *goal-oriented behavior* as consisting of *motivated choices* made by individual *agents*. These choices are goal-oriented in that *actions* are *selected* based on the degree to which agents *differentially value* (i.e., *prefer*) particular outcomes, as well as *beliefs* about the relative likelihoods with which these outcomes will result from particular actions. EDT considers agents to be *"rational"* to the extent that choices maximize *overall value* (i.e., *utility*). Choice utility can be estimated by *expected value analyses* that integrate the desirability of all possible outcomes, weighted by the respective probabilities with which they are likely to occur.

For the sake of illustration, let us consider a human-like agent choosing between apples or oranges to satiate feelings of hunger. In addition to ascribing value to actions where outcomes alleviate the *pain* of hunger, the agent believes that the amount of *pleasure* it experiences will be determined by the manner in which these *desires* are satiated. That is, the agent has specific beliefs about the *value* of *anticipated experiences resulting from different choices* (i.e., *expected utility*), the accuracy of which will depend on the *actual* *experiences associated with choices* (i.e., *experienced utility*). Considering that different pieces of fruit vary in quality, there is also uncertainty in the expectation that a particular selection will maximize experienced utility. In this way, each fruit selection represents a kind of lottery where identical choices can result in discrepant outcomes. Nonetheless, even though anticipated pleasure and pain may not coincide with actual experience in any given instance, a rational agent will choose depending on the *expected utility associated with different choices*.

More specifically, agents typically value increased pleasure or decreased pain as having *positive expected utility*, and devalue increased pain or decreased pleasure as having *negative expected utility*. For example, if an agent previously experienced greater pleasure increases from apples compared to oranges, then it may assign a higher positive expected utility to apple selections. Alternatively, if an agent has previously experienced greater pain increases with oranges compared to apples—e.g. experiencing an allergic reaction—then it may assign a higher negative expected utility to orange selections. In either of these cases, a hungry and rational agent would select foods where outcomes are expected to maximize overall pleasure and minimize overall pain.

These *differential expected utilities* (i.e., preferences) can be represented by abstract mathematical expressions known as *utility functions*, which specify relative valuations of different outcomes. Theoretically, these functions correspond to likelihoods with which rational agents will make various choices in situations involving preferred outcomes. Utility functions can be *estimated* from both explicitly stated preferences—to the degree that subjective reports are accurate—as well as *from observed choices* (i.e., *revealed preferences*) (Samuelson, 1938). However, as with all models, these empirically derived utility functions only imperfectly estimate the *unobservable variables* (i.e., *latent variables*) of an agent's actual preferences. Additionally, these approximations are necessarily limited to the extent that revealed preferences are incomplete measures of the overall value systems determining expected utilities. That is, since choices have different consequences in different circumstances, the magnitude of individual preferences can depend on an indeterminate number of context-specific factors.

With respect to our hungry agent, its preferences for apples and oranges in varying circumstances could be specified in a utility function for relative fruit valuations. To the extent that the agent is rational, this function would predict the likelihoods of its choices in different situations, and could be estimated from stated food preferences as well as food-selecting behaviors. However, even under an extremely simple analysis only concerned with present enjoyment, foods may be differentially valued in different combinations, such that preferences revealed by particular actions or statements might not provide an accurate description of its actual utility function. Although there may be a simple relationship describing the value assigned to each additional apple or orange, the idiosyncratic preferences of an agent may be far more complicated. Further, the value of eating a particular kind of food may vary with needs for calories and nutrients, anticipation of the availability of nutrition in the future, concerns for health and appearance, or numerous other factors that could impact expected utility. Over-simplified utility functions can be useful for modeling preferences within limited domains, but in order to predict food choices accurately, these different kinds of value must all be considered. Unfortunately for economists, this goal is unattainable in light of the unbounded complexity of potential value systems.

## BOUNDED RATIONALITY AND SUBOPTIMAL CHOICES

*"The probability of any event is the ratio between the value at which an expectation depending on the happening of the event ought to be computed, and the value of the thing expected upon its happening."*
        –Thomas Bayes (1701-1761)

The imperfect rationality of agents further limits the reliability of estimated utility functions. That is, although optimal utility maximization may be desired, agents have necessarily limited accuracy when inferring the actual likelihoods of relevant action-outcome associations. Thus, agents necessarily exhibit *bounded rationality* in that they deviate from the unattainable ideal of integrating all relevant information into a single probabilistic model (B. D. Jones, 1999; Simon, 1955). Further obstacles to rational optimality include reasoning errors from cognitive limitations, reliance upon imperfect heuristics, and emotion-driven biases (Daniel Kahneman, 2003, 2011; Daniel Kahneman & Tversky, 2007; Simon, Egidi, Viale, & Marris, 2008).

To return to our hungry agent, bounded rationality can compromise utility maximization in numerous and varied ways. For example, the agent may not realize that a particular batch of oranges was grown under optimal conditions for maximizing nutrition and flavor. Alternatively, it may think that apple-growing conditions were ideal, when in fact the high humidity was also ideal for a particular species of apple-infesting fungus. Further, the agent may not realize that it has a vitamin C deficiency that will lead to health problems if not remedied, or that pesticide residues on apple skins increase the risk of diseases to which the agent may be particularly vulnerable. The agent may fail to integrate these factors because of a lack of available information, insufficient mental capacity, or perhaps some sort of irrational prejudice against oranges or bias in favor of apples. Thus, multiple factors can contribute to our agent being mistaken in its beliefs about the expected utilities of various choices and their associated outcomes.

If choices are compromised by ignorance or irrationality, then empirically-derived utility functions will inaccurately model value-consistent preferences to the extent that they do not account for these factors. Although such considerations limit the ability of economists to predict choices, attempts to more accurately model human action in behavioral economics are still grounded in EDT as a kind of default null-hypothesis (D. Kahneman & Tversky, 1979; Trepel, Fox, & Poldrack, 2005). That is, even when theories emphasize the bounds of reason, these models are still expressed in terms of deviations from optimal utility maximization.

## ACTION SELECTION, EXPECTATIONS, AND EVOLVING PREFERENCES

*"The consequences of an act affect the probability of it's occurring again."*
        –B.F. Skinner (1904-1990)

For biological agents, choices emerge from underlying patterns of neural activity that ultimately result in specific motor sequences. However, since actions must be coordinated in order to produce coherent behavior and achieve particular outcomes, there is a sense in which corresponding neural representations compete and cooperate in determining which patterns will be expressed in which order within hierarchies of actions and sub-actions. *Action selection*—the process of arbitrating among interacting behavioral representations in determining what to do next—remains a central topic of investigation for neuroscience, as well as in the fields of artificial intelligence and ethology (Gurney, Prescott, & Redgrave, 2001; Houk et al., 2007; Humphries, Gurney, & Prescott, 2007).

For example, if our hungry agent encounters apple and orange trees in the same location, concurrently elevated levels of neural activity may occur for representations involved in obtaining and consuming these various fruits. However, since these patterns share a common embodiment, they cannot simultaneously direct effector systems without compromising food obtainment (e.g. attempting to climb both trees at the same time). In any given moment, neural representations for specific actions need to be prioritized (i.e., selected) depending on which outcomes are more likely to maximize utility. In this way, the processes involved in selecting actions are continuous with those involved in preferring them. Subjectively, the magnitude of differential expected utility corresponds to the strength with which particular outcomes are preferred, as well as the degree of motivation for realizing these desires.

In terms of motivated behavior, outcomes are valued as *rewarding* or devalued as *punishing* if they respectively increase or decrease overall expected utility. That is, rational agents will be more likely to select options they expect to increase rewarding experiences and decrease punishing experiences, thereby optimizing choices for utility maximization. In an operant conditioning framework (Skinner, 1938), the impacts of outcomes are deemed *reinforcers* or *punishers* depending on whether they respectively increase or decrease the likelihood of behaviors. Reinforcement and punishment—as well as their influences on action-tendencies—can occur with or without conscious awareness (Both et al., 2008; Hoffmann, Janssen, & Turner, 2004; Morris, Ohman, & Dolan, 1998). The potential for unconscious processes to impact behavior further limits the reliability of utility functions based on explicitly stated preferences. Nonetheless, the principles of EDT remain useful for describing choices, even if driven by processes lacking awareness or intentionality.

From a MEDO perspective, specific preferences—as well as associated beliefs and information processes—probabilistically influence which actions and reactions are likely to occur in various situations. To the extent that different goals constitute independent sources of value, they can be expressed as separate terms in a utility function describing overall motivation (Abbas, 2010; D. E. Bell, 1979; Gw, W, D, & M, 1995). With respect to our hungry agent, its independent valuations of apples and oranges could be expressed as separate terms in its utility function for food preferences. The relative weightings of these terms would specify the *magnitudes of positive and negative expected utilities* (i.e., *reward* and *punishment expectations*, respectively) associated with consuming these various foods. Moreover, since both expected rewards and punishments can contribute to differential expected utilities (i.e., preferences), both potential increases and decreases in value can drive motivation and desire.

However, since experiences shape expected utility, relative valuations are not static quantities. Expected utilities increase or decrease based on the number, intensity, and kinds of rewarding and punishing experiences. Hence, additional experience could further change valuations. With respect to the abstract utility function describing these preferences, corresponding terms would also increase or decrease to reflect these changing significances. For example, if reward expectations for oranges increase after encountering a particularly appetizing batch, then relative preferences for apples would necessarily decrease. Even with constant expectations for consummatory pleasure, enhanced desire for oranges would necessarily decrease the proportional expected contributions of apple selections to overall utility. Conversely, if particularly unappetizing experiences increase expected punishment from apples, then this revaluation of alternative choices would increase relative preferences for oranges. In either of these cases, preferring one option more entails preferring the other option less, and vice versa. Alternatively, apples and oranges could be especially delicious in combination, thereby increasing the positive expected utilities of both fruits without changing expected utilities relative to each other. Nonetheless, all preferences are necessarily relative as long as actions are constrained by limited resources.

## FEEDBACK DYNAMICS AND DIFFERENTIAL AMPLIFICATION OF PREFERENCES

*"Positive feedback loops are sources of growth, explosion, erosion, and collapse in systems. A system with an unchecked positive loop ultimately will destroy itself. That's why there are so few of them. Usually a negative loop will kick in sooner or later."*
        –Donella Medows (1941-2001)

By influencing action selection, changing valuations can shape behavior in ways that further alter preferences. To the extent that likelihoods for high-quality fruit selections can increase through practice, then reward expectations would also increase with picking skill. However, since likelihoods for obtaining this practice are influenced by the ability of expected rewards to bias action selection, this *feedback loop* could amplify reward expectations with time and experience. This sort of dynamic could also be supported by progressively decreasing punishment expectations, such as reducing likelihoods of selecting low-quality fruit or difficulties with orange peeling. Theoretically, a sufficient number of positive experiences could result in oranges being exclusively preferred, thereby eliminating the expected utility of apples.

This example involved *positive feedback*, which refers to a process increasing as a result of its outputs. In this case, relative preferences drove action selection (i.e., the process) such that experienced utility further changed relative expected rewards (i.e., the outputs), thereby resulting in increasingly amplified differential expected utility. *Negative feedback* corresponds to the opposite situation of a process decreasing as a result of its outputs, such as would be the case if increased experiences with a kind of fruit resulted in boredom, thereby reducing expected rewards. Yet the resulting attenuation of expected utility would be *self-limiting*, as less frequent selections would reduce boredom.

Punishing experiences could influence relative preferences via negative feedback, since fruit with higher punishment expectations would be less likely to be selected. This process of increasing negative expected utility for punished fruit selections would also be self-limiting. However, this bias could further contribute to the positive feedback dynamics of increasing relative preferences for non-punished fruit selections. Theoretically, intensely punishing early experiences with apples could result in an agent always choosing oranges, even if these initial apples were of unrepresentatively poor quality. In any of these cases, if an agent lacks motivation for choosing apples, then this relative devaluation would be perpetuated by self-sustaining patterns of goal-oriented behaviors. Alternatively, if apples and oranges were synergistically rewarding in combination, then rather than feedback amplification of initial differences in expected value, expected rewards for apples could increase as a consequence of increasing skill in orange selection, and vice versa.

Thus, not only do neural representations for expected utilities compete and cooperate in influencing particular choices, but experience-driven expectations also allow this evolution to be extended over longer timescales through feedback dynamics. *Differential feedback amplification of reward and punishment expectations* could lead to action selection being dominated by one preference or another, thus resulting in *increasing specialization and differentiation in a particular direction.* It is also possible to have *cooperative synergy among expectations*, but all preferences have associated *opportunity costs with respect to other potential dynamics of differential specialization* (e.g. developing skill in picking or preparing other types of fruits). In this way, *each preference competes and cooperates with every other preference in determining the future evolution of the overall system*.

## GENERALIZED DARWINISM

### NATURAL SELECTION: REPLICATION, VARIATION, SELECTION

*"The generality of the principles of natural selection means that any entities in nature that have variation, reproduction, and heritability may evolve... This axiomatization makes clear that the principles can be applied equally to genes, organisms, populations, species, and at opposite ends of the scale, prebiotic molecules and ecosystems."*
      –Richard Lewontin (1929-present)

MEDO analyzes *preferences as action selection biases*, which influence the ability of various neural patterns to shape behavior. However, as with all other biological patterns, the neural representations underlying these *preferences are also modeled as competing and cooperating in a Darwinian struggle for existence* (Aldrich et al., 2008; Edelman, 1993; Edelman & Mountcastle, 1978; Hayek, 1952; McDowell, 2010; Seth & Baars, 2005; Wyckoff, 1987).[1] MEDO further considers *selective processes to be implicit—and sometimes explicit—preferences emerging from evolutionary systems*. From this perspective, EDT can be used to characterize both the motivations of individual agents, as well as dynamics within evolutionary systems more generally. In this way, EDT provides a common language for modeling selection and preferences on multiple levels, including population genetics, cultures, and within individual nervous systems.

---

[1] Neural Darwinism will be described in greater detail in the section "Phenotypic plasticity and the evolution of evolving nervous systems."

MEDO utilizes evolution as a general principle for understanding the emergence of complexity within all dynamical systems capable of supporting *replication*, *variation*, and *selection*, regardless of specific details of mechanistic implementation (Aldrich et al., 2008; Campbell, 2009; R. Dawkins, 1983; Nelson, 2007). These three features are the *necessary and sufficient conditions for natural selection* to occur:

- Replication: *replicators* are any patterns capable of reproducing aspects of themselves over time, and thus exhibiting some degree of heredity—broadly construed—in their *ability to increase the likelihood that similar forms will persist into the future* (i.e., *evolutionary fitness*).
- Variation: *variations* are features that differ between replicators and *influence relative evolutionary fitness* (i.e., *adaptive significance*).
- Selection: *selective pressures* are any factors capable of influencing adaptive significance.

These principles are most frequently associated with biological species. More specifically, adaptive significance applies to the *sum-total of biological features* (i.e., *phenotypes*) within populations of organisms. Further, evolutionary fitness applies to the replicative success of heritable aspects of organisms (i.e., *genes)* contributing to these phenotypes.

However, evolution is not limited to reproducing organisms and population genetics. Indeed, when Williams (1966) first suggested the *gene-centered view of evolution*, he defined a gene in probabilistic terms as "*that which segregates and recombines with appreciable frequency*" (Williams, 1966). Darwin (1859) himself proposed that the principles of natural selection could apply to the formation and transformations of languages over time, as well as competing customs among tribal groups (Aldrich et al., 2008). In a similar vein, Dawkins (1976) suggested that aspects of cultural evolution can be understood in terms of "memes," or units of information that compete and cooperate to perpetuate themselves within and between the minds of humans (Richard Dawkins, 1976).[2] Further, evolutionary history preceded the advent of DNA-based inheritance, beginning with self-perpetuating cycles of chemical reactions and possibly self-replicating RNA molecules (Takeuchi, Hogeweg, & Koonin, 2011; Takeuchi, Salazar, Poole, & Hogeweg, 2008). Moreover, even self-modifying proteins and computer programs are capable of evolving by natural selection (Bawazer et al., 2012; Holland, 1992; Shorter, 2010; Von Neumann, 1966; Waters, 2011). Thus, Darwinian analysis can extend to a broad range of phenomena, including the competition and cooperation of neural patterns within the minds of individual agents.

---

[2] Numerous criticisms have been levied against the meme of memes, ranging from the analogy being misleading on account of dissimilarities with genetic replicators, to the memetic evolution providing an inaccurately over-simplified account of cultural change (Atran, 2001; Boyd & Richerson, 2000; Fracchia & Lewontin, 2005).

## NATURAL SELECTION, ADAPTATION, AND EVOLUTIONARY UTILITY

*"There cannot be design without a designer; contrivance without a contriver; order without choice; arrangement without anything capable of arranging; subservience and relation to a purpose without that which could intend a purpose; means suitable to an end, and executing their office in accomplishing that end without ever having been contemplated or the means accommodated to it. Arrangement, disposition of parts, subservience of means to an end, relation of instruments to a use imply the presence of intelligence and mind."*
　　–William Paley (1743-1805)

*"Adaptation and function are two aspects of one problem."*
　　–Julian Huxley (1887-1975)

Most simply, natural selection can be viewed as an elaboration of a seemingly tautological maxim: *patterns that replicate best tend to replicate most.* That is, if heritable variations are more effective at increasing replicative success, then they are more likely to predominate within a system. However, when selected change is able to accumulate over geological timescales—as with genetic heredity—this simple heuristic explains how intricately functioning biological systems can arise from "blind" processes lacking conscious foresight or intentionality (Richard Dawkins, 1996a, 1997; Dennett, 1996, 2009). Thus, evolution creates a kind of *illusory goal-directedness*, in that resulting forms appear to be intentionally *designed* (i.e., *engineered*) in their *usefully specialized complexity* (i.e., *adaptations*). However, these complex functional specializations emerge from the simple fact that some patterns are more effective at replicating than others. In this way, evolution allows for "*design without a designer*" (Ayala, 2007), and adaptations can be understood as *partially optimized solutions for engineering problems* (Dennett, 1996).

For example, if apples and oranges were plentiful in an environment with scarce nutrition, then organisms would be more likely to survive and reproduce if they were better suited for achieving sustenance from these foods. That is, to obtain nutrients from these sources, various characteristics of different organisms would make them more or less effective at acquiring fruits, identifying particularly edible specimens, extracting the most palatable parts, etc. Each of these challenges would constitute selective pressures in that an organism's ability to perform these functions would influence its ability to reproduce (i.e., evolutionary fitness). To the extent that these features were heritable, with subsequent generations, their prevalence in the overall population would increase as a result of their positive adaptive significance. On sufficiently extended timescales, the average member of that species—or sub-species—could eventually acquire complex functional specializations (i.e., adaptations) for effectively achieving sustenance from apples and oranges.

In this way, selective pressures emerge from interactions between organisms and *frequently encountered environments*, also known as *ecological niches* (Begon, Harper, & Townsend, 1999). More specifically, the relationships between organisms and their niches are primary determinants of selection, since these interactions influence the survival and reproductive consequences of various adaptations. Over time, species tend to become increasingly well adapted to their respective niches, since these environmental conditions influence which organisms produce more descendants. However, through their functioning, organisms are also capable of altering environments in a process known as *niche construction* (Day, Laland, & Odling-Smee, 2003; K N Laland, Odling-Smee, & Feldman, 1999; Odling-Smee, Laland, & Feldman, 2003). This bidirectional flow of information between organisms and their environments occurs on the level of individuals, groups, and even entire species. Over time, these interactions are capable of radically changing environmental conditions, thereby altering the selective pressures that initially shaped them.

There are numerous ways in which niche construction could impact our population of evolving fruit eaters. For instance, a coalition of individuals could focus on a particular expanse of trees and collectively defend it against other organisms. In this case, the fitness of these individuals would be primarily determined by their functioning within this territory, as well as by their interactions with other group members. Additionally, if these organisms learned how to select apples and oranges with optimal nutritional properties, then the seeds of those particular fruits could be dispersed and fertilized with greater frequency. Over time, this could lead to a greater proportion of trees producing highly nutritious apples and oranges, which in turn could lead to those fruits producing increasing reward expectancies in the animals that consume them.

Theoretically, an initial discrepancy in the preferences of fruit consuming—and seed distributing—organisms could result in differential feedback amplification of co-evolving species being greater for one variety of fruit compared with another. If this situation resulted in an environment dominated by orange trees, then this cooperative evolution would come at the expense of organisms that were adapted to apple trees as their primary niche. Alternatively, if apple-dependent organisms were necessary for maintaining ecosystem stability, then differential feedback amplification could be self-limiting via a negative feedback relationship between orange dominated environments and the conditions supporting this proliferation.

Thus, context-dependent changes in adaptive significances allow natural selection and niche construction to be functionally interrelated through processes of *reciprocal causation* (i.e., *bidirectional causation* or *circular causation*), thereby potentially enabling both enhancement and inhibition of specialization and differentiation in populations of organisms. Notably, reciprocally causal relationships also characterize our previously described example, wherein experience-dependent changes in expected utilities allowed preference-based action selection and practice-based skill learning to have bidirectional influences with each other. Although one case describes goal-oriented behavior, and the other purposeless optimization, both scenarios involve *self-modifying evolution determined by various positive and negative feedback processes on multiple levels*. Additionally, in both cases, the balance between these feedback dynamics largely depends on happenstance, in that defining contingencies could be influenced by chance events such as idiosyncratic historical circumstances (P Bak & Paczuski, 1995; Per Bak, Tang, & Wiesenfeld, 1987; Lovecchio, Allegrini, Geneston, West, & Grigolini, 2012).

While discussing the lack of ultimate purpose in evolution, Dawkins (1996) proposed that natural selection can be expressed in terms of an *evolutionary "utility function"* in which value solely depends upon genetic proliferation (Richard Dawkins, 1996b). Even if adaptations result from illusory goal-directedness, this analogy appropriately uses the language of EDT. *Functionality entails relative optimality, and hence utility.* However, the rationality of mindless evolution is bounded in that changes can only be made through incremental modifications of existing adaptations, rather than explicitly planned designs. Since evolution cannot be informed by expectations of future environmental changes, neither the processes involved nor outcomes produced are guaranteed to be optimal in any sense.

For example, our fruit-eating organisms could theoretically extract more nutrients if they were capable of metabolizing entire oranges. However, even if this functionality could be achieved with a simple enzyme for digesting peels, requisite adaptive variations may have never arisen over the evolutionary history of that species. Or, at one point in time, there may have been organisms capable of digesting peels, but selective pressures may have been insufficiently powerful to reliably increase the prevalence of those adaptations. In natural selection, particular forms come to predominate because they were better than alternatives *under the conditions in which they were selected*, also known as the *environment of evolutionary adaptation* (Foley, 1995; Segal & MacDonald, 1998; Tooby & Cosmides, 1990). However, better in no way implies that particular adaptations are the best conceivable solutions for their respective functions (Richard Dawkins, 1997; Wright, 1982).

## NATURAL SELECTION, ADAPTIVE SIGNIFICANCE, AND EVOLVING SELECTIVE PRESSURES

*"Since the phenotype as a whole is the target of selection, it is impossible to improve simultaneously all aspects of the phenotype to the same degree."*
    –Ernst Mayr (1904-2005)

MEDO applies EDT to all evolutionary systems, even if they lack explicit goals or beliefs required for intentional behavior. Although only agents have subjective experiences, *non-agent systems can still be considered to have implicit preferences in that certain outcomes (i.e., variations) are favored over others*. That is, as replicators compete and cooperate in attempting to perpetuate themselves with limited resources, variations are differentially selected to the extent that they effectively satisfy fitness criteria (Richard Dawkins, 1997; Wright, 1932). Or more precisely, environments do not actively select particular forms, but the most effective replicators will tend to predominate within the systems in which they perpetuate themselves. Nonetheless, *selective pressures can be viewed as terms in an abstract utility function for evolutionary fitness*: just as the varying preferences of agents compete and cooperate in influencing choices, every selective pressure competes and cooperates with every other selective pressure in influencing adaptive significance.

Mindless selective processes are incapable of conducting expected value analyses to maximize their implicit preferences.[3] However, this sort of computation is automatically carried out by the numerous and varied interactions of organisms with their environments, given sufficiently large populations. These interactions—and their resulting impacts on differential replication—implement a kind of sampling process that incorporates both adaptive significances of outcomes as well as probabilities for their occurrences in environments of evolutionary adaptation. This process may lack intentionality or beliefs, yet *to the extent that the future resembles the past, prior evolutionary outcomes can function analogously to accurate reward and punishment expectancies* for the motivated behavior of individual agents. To the extent that selective pressures change over time, then differential survival and reproduction may eventually alter population characteristics in a way that reflects these changing adaptive significances. That is, novel selective pressures can update evolutionary outcomes, given that the population survives. Alternatively, entire classes of organisms can go extinct, as has been the case for the vast majority of species throughout evolutionary history.

To return to our example of evolving fruit preferences, if apples and oranges had synergistic nutritional properties when combined, then selection could increase for adaptations relating to both foods. However, if fitness depended upon oranges more than apples in the environment of evolutionary adaptation, then it can also be expected that organisms would become particularly well suited to obtaining nutrition from oranges, relative to apples. With respect to the implicit utility function describing these selective pressures, terms corresponding to the adaptive significances of orange-related specializations would be more heavily weighted, relative to terms describing apple-related specializations.

Theoretically, the average organism within the population could become so thoroughly adapted to eating oranges that adaptive significance lessened for apple-related specializations. This scenario would be especially likely if functional tradeoffs were required in optimizing animals for unique properties differing between apples and oranges. For example, if more minerals were deposited into strong jaws for consuming apples, then fewer minerals might be available to strengthen fingers for peeling oranges. Over time, declining selective pressures could result in a situation where apple-related specializations deteriorate to the point of non-functionality (i.e., *vestigiality*), thus completely eliminating contributions from apple-related adaptations to evolutionary fitness. Apple-specific nutrients could become essential for survival in future environments, but the population would be unable to respond to these challenges if it no longer contained sufficient adaptations for consuming apples. In this scenario, the bounded rationality of evolution might result in extinction. However, if apple-related adaptations remained in a subset of organisms, then changing selective pressures could result in those features spreading throughout the population. Additionally, functional tradeoffs might result in organisms becoming decreasingly effective at obtaining sustenance from oranges, depending on the specific feedback dynamics involved.

---

[3] Although evolution does not use expected value analyses to determine preferences, neither do most humans. The complexity of potential consequences produces a combinatorial explosion, necessitating sampling in exploring these probability spaces. However, unlike blind natural selection, human minds can use causal models of future events to restrict searches. Nonetheless, since modeled futures are influenced by the limitations of past experience, even conscious beings can be impeded in maximizing value if they become fixated on suboptimal solutions.

As this thought experiment reveals, in addition to selective pressures competing and cooperating in shaping dynamics, evolutionary outcomes also compete and cooperate through their relative contributions to overall fitness. Just as with agent choices, differential feedback amplification can produce specialization and differentiation within evolutionary systems. For agents, past expected utilities influence present outcomes, which in turn contribute to current preferences that influence present and future expected utilities. For evolutionary systems, past selective pressures influence present outcomes, which in turn contribute to current selective pressures that influence present and future adaptive significances. Although these feedback dynamics involve conscious agents in one instance and unconscious optimization processes in the other, MEDO uses common formalisms to analyze underlying processes.

## OPTIMIZATION ON MULTIPLE LEVELS

### MULTI-OBJECTIVE OPTIMIZATION PROBLEMS AND EVOLUTIONARY ENGINEERING ANALYSES

*"It is a profound truth that Nature does not know best; that genetical evolution... is a story of waste, makeshift, compromise and blunder."*
        –Peter Medawar (1915-1987)

Although adaptations result from blind optimization processes, the complex functional specializations of biological systems can nonetheless be analyzed with the same principles used for engineering decisions (Lewis et al., 2006). Indeed, the dynamics underlying evolutionary adaptation have similar formal properties to those found in *multi-objective optimization problems* (Coello et al., 2009; Keeney & Raiffa, 1993; Sawaragi, Nakayama, & Tanino, 1985). That is, each attribute of a design represents a separate source of expected utility in contributing to overall value.

More specifically, in choosing between various design configurations, engineers decide which *functional tradeoffs* (e.g. optimizing for affordability *or* reliability) and *functional synergies* (e.g. optimizing for reliability *and* durability) will be preferred in attempting to maximize the overall utility of their creations. Just as the neural representations for expected utilities compete and cooperate in shaping action selection, engineering always involves attempting to satisfy multiple objectives with minimal compromises—even when this "engineering" is carried out by purposeless natural selection (e.g. optimizing organisms for reproductive capacity *or* robustness (Pianka, 1970; Reznick, Bryant, & Bashey, 2002)). Thus, the principles of EDT can be logically extended to all optimization processes, whether goal-directedness is illusory or actual.

In the same way that observed behavior can be used to estimate relative preferences, by viewing the *functional properties of designs as choices* (i.e., revealed preferences), it is possible to estimate sources of utility influencing optimal solutions. For example, if a product automatically peels oranges, then it could reasonably be inferred that the engineer preferred for the device to have this functionality. It could be additionally inferred that this feature was expected to either directly provide value for the designer—who may have had a particular fondness for eating oranges or a particular aversion to peeling them—or indirectly by providing value to potential customers. If the machine incompletely peels the oranges it encounters, then it could also be inferred that this outcome was desired. Perhaps there was a functional tradeoff where complete peeling could only be achieved by greatly extending operating times, which would have been frustrating for both the designer and purchasers. Or, creating a more thorough peeler could have increased costs to a point where the device was prohibitively expensive for potential customers. In either of these two cases, design characteristics would reflect the competing and cooperating sources of expected utilities motivating engineering decisions.

However, just as utility functions inferred from agent choices have limited reliability, similar limitations apply when attempting to infer relative expected utilities from engineered designs. In other words, all agents make suboptimal decisions, including engineers. With respect to the aforementioned ineffective orange peeler, the designer may have had a personal preference for orange rinds, and incorrectly assumed that the majority of consumers feel similarly. If this inaccurate belief resulted in reduced sales—and if sales were valued—then bounded rationality lead to actions whose outcomes did not maximize utility. Or, the engineer may have desired a thorough orange peeler, but lacked the capability to design such a product. In either of these two cases, some of the engineer's relative preferences would be incorrectly inferred from design characteristics.

Nonetheless, just as agents differ in their degrees of rationality, engineers differ in their abilities to design systems according to their desires. The greater the skill of an engineer, the more likely it is that design characteristics will represent intentional choices, thereby reflecting the utility function of the optimizer. If an engineering genius made a product that consistently under-peels oranges by a slight amount, then that particular outcome is likely to be precisely what maximized overall utility. That is, the outcome maximized value given the necessary constraints of functional tradeoffs and synergies, which determine the competitive and cooperative relationships between various preferences. Assuming a high degree of skill and accurate beliefs regarding action-outcome associations, *relative strengths of preferences can be expected to proportionally drive specific design choices* for a rational engineer.

In light of the fact that evolution always involves multiple selective pressures, similar considerations apply to the emergent designs of natural selection. Just as the skill of an engineer determines the degree to which preferences can be inferred from design characteristics, the power of an evolutionary optimization process determines the degree to which selective pressures can be inferred from impacted phenomena. Thus, *viewing evolutionary outcomes (i.e., adaptations and their frequencies within populations) as* *signifying past selective pressures is comparable to viewing the outcomes of agent choices as revealed preferences.*

However, if the power of natural selection is compromised by historical circumstance or chance events (Eldredge & Gould, 1972; Gould, 1990), then empirically derived selective pressures will inaccurately model past fitness criteria to the extent that they do not account for these factors. Although such considerations limit the ability of biologists to infer adaptive significances, attempts to more accurately model evolution are still grounded in an engineering perspective as a kind of default null-hypothesis (Gould, 1992). That is, even when theories emphasize the bounds of evolutionary optimization, these models are still expressed in terms of deviations from optimal fitness maximization (Gould & Lewontin, 1979).

## EVOLUTIONARY COMPUTATION AND FITNESS LANDSCAPES

*"In the theory with which we have to deal, Absolute Ignorance is the artificer; so that we may enunciate as the fundamental principle of the whole system, that, in order to make a perfect and beautiful machine, it is not requisite to know how to make it. This proposition will be found, on careful examination, to express, in condensed form, the essential purport of the Theory, and to express in a few words all Mr. Darwin's meaning; who, by a strange inversion of reasoning, seems to think Absolute Ignorance fully qualified to take the place of Absolute Wisdom in all of the achievements of creative skill."*
    – Robert Mackenzie Beverley (1798-1868)

*"Everyone by now presumably knows about the danger of premature optimization. I think we should be just as worried about premature design - designing too early what a program should do."*
    –Paul Graham (1964-present)

E*volutionary computation* combines aspects of intentional and unintentional design in ways that illustrate key principles of the MEDO framework (Coello et al., 2009; Keeney & Raiffa, 1993; Sawaragi et al., 1985). With this approach to optimization, methodologies such as *genetic algorithms* are commonly used to solve multi-objective problems for which design considerations are too complicated for intentional engineering.[4] However, with these techniques, engineers can *indirectly maximize utility by defining selection criteria* (i.e., *fitness functions*) for programs that evaluate various design configurations. These algorithms search through large numbers of feature permutations—requiring super-computing to run these simulations—and probabilistically choose the subset of designs that best meet pre-specified selection criteria, thereby maximizing their fitness functions. Alterations are introduced in the process of replicating successful configurations, thus creating *variations* in the next generation of designs from which further selection occurs.

---

[4] Although evolutionary computation is a broad discipline involving many techniques, we will focus on genetic algorithms because of their conceptual similarities with core principles of biological natural selection. Also, we will only discuss a subset of the many different ways that genetic algorithms are utilized to solve optimization problems.

With subsequent iterations, increasingly well-adapted solutions can be discovered for complex engineering problems.

For example, an engineer may desire to create affordable, reliable, quick, neat, and thorough orange peelers, but is unable to determine which combination of features and materials would be most likely to maximize sales while minimizing production costs. Theoretically, our aspiring entrepreneur could use genetic algorithms to attempt to solve this problem, given that sufficient funds are available to pay a programmer and obtain processing time on a super-computing cluster. Rather than designing the product directly, the programmer would design the simulation environment and algorithms, and the engineer could specify selection criteria that maximize profits while incorporating these various considerations, including functional tradeoffs and synergies.

If skillfully implemented, genetic algorithms produce increasingly optimized solutions with successive generations. However, just as with biological evolution, these programs are not guaranteed to converge upon optimal solutions. Proposals will be better than alternatives, but not necessarily best overall. In our example, the algorithm could select increasingly useful and profitable designs over time, yet they could all be inferior to the engineer's initial orange peelers. Additionally, it is a non-trivial challenge to choose fitness functions that select design configurations with high utility (e.g. overall profits). That is, the fitness function of an algorithm may not correspond to the utility function of the engineer. This may be due to the fact that the programmer is unable to create a simulation environment that adequately models design challenges, or selects configurations in ways that inaccurately represent engineering preferences. Alternatively, the principle of "garbage in, garbage out" could apply if specific preferences would be unlikely to generate utility if realized outside of computer simulations. In this case, the bounded rationality of the engineer—rather than the programmer—would be responsible for suboptimal choices.

The efficacy of genetic algorithms can be visualized using *fitness landscapes*, where base dimensions are defined by various features, and graph-heights represent evaluations of fitness functions with different combinations of feature specifications (Richter, 2010). This conceptual approach was first proposed for evaluating the evolutionary fitness of various genotypes and phenotypes, where similar configurations are located closer to each other on landscapes, relative to dissimilar configurations (Richard Dawkins, 1997; A. J. Eckert & Dyer, 2012; Misevic, Kouyos, & Bonhoeffer, 2009; Wright, 1932). For genetic algorithms, a given set of designs can be represented as distributed points on a landscape, with better solutions located higher on the graph, relative to worse solutions. When random variations are applied to previous configurations with greatest fitness, some of these configurations will increase optimality, and some will decrease optimality. That is, random lateral shifts on a fitness landscape will cause solutions to move to higher elevations, and some to move to lower elevations. If the algorithm is more likely to select configurations with greater fitness, then populations of candidate solutions will climb to higher elevations with successive generations, such that the average design configuration is increasingly likely to be distributed near fitness peaks.

The overall efficacy of genetic algorithms can be conceptualized as the reliability and speed with which proposed solutions reach *landscape peaks* (i.e., *maxima*), corresponding to *locally optimal solutions*. However, since solutions with lower elevations are less likely to be selected, genetic algorithms can become trapped at local optima and fail to reach *globally optimal solutions* during the time allotted for simulation. More specifically, variability is typically introduced into these programs as minor alterations, corresponding to small lateral shifts on a fitness landscape. This search strategy is justified based on the notion that a series of small shifts is more likely to result in progressive movement to higher elevations, relative to large shifts that may move a design configuration away from peaks. However, if relatively optimal candidate designs are located on local optima, they are unlikely to move to other peaks if it requires a series of moves to lower elevations. Configurations closer to the local maxima may be created in any given generation, but if these shifts involve moving to lower elevations, they are less likely to be selected for the next generation. In this way, the majority of candidate solutions are likely to return back to local maxima with subsequent generations.

Genetic algorithms can be modified to become more effective at maximizing overall fitness. Yet, this usually depends on a conscious engineer-programmer who evaluates outcomes and explicitly decides to change fitness parameters (Smit & Eiben, 2009). In our example, the engineer and programmer could attempt to simplify the problem by removing operating-time as a consideration, thereby *reducing the number of objectives that need to be optimized*. Alternatively, they could decide to prioritize neatness or thoroughness to greater or lesser degrees—thus, *changing weighting coefficients of different objectives in the fitness function*—depending on the extent to which these various alterations produce desirable outcomes. In terms of fitness landscapes, either of these strategies could *reduce the variability of the landscape, thereby potentially increasing the likelihood that algorithms reach global optima.*

## INTENTIONALITY AND EVOLVING UTILITY LANDSCAPES

*"Probably the first clear insight into the deep nature of control… was that it is not about pulling levers to produce intended and inexorable results. This notion of control applies only to trivial machines. It never applies to a total system that includes any kind of probabilistic element—from the weather, to people; from markets, to the political economy. No: the characteristic of a non-trivial system that is under control, is that despite dealing with variables too many to count, too uncertain to express, and too difficult even to understand, something can be done to generate a predictable goal."*
     –Anthony Stafford Beer (1926-2002)

Optimality notwithstanding, by specifying fitness functions for genetic algorithms, conscious agents can potentially maximize value from processes lacking foresight or intentionality. Similarly, *MEDO stipulates that genes maximized replicative success by influencing—with limited specificity and bounded rationality—fitness criteria for evolutionary optimization within the nervous systems of individual organisms*. Yet, in this case, a blind optimization process eventually gave rise to something new in the history of evolution: self-aware minds in which explicit models could intentionally drive goal-oriented behavior. Even without self-awareness, the utility functions of biological agents continuously change with experience in ways that increase the efficacy with which they achieve valued outcomes (i.e., operant conditioning). Similarly to how engineers could update fitness functions to increase the efficacy of genetic algorithms, agents can update their preferences to increase their ability to maximize overall utility.

However, the manner in which engineers indirectly maximize utility by altering computer algorithms is more closely paralleled by the manner in which genes indirectly maximize fitness by shaping the development of nervous systems. More specifically, *while genes can only determine adaptive significance from post hoc consequences, self-aware agents can continuously alter their particular expectations and preferences based on moment-to-moment estimations of dynamics*. That is, explicit models allow organisms to more precisely anticipate the consequences of their choices in novel situations before acting upon them. This increased predictive power allows organisms to more effectively function within their various niches and discover fitness maxima. However, this flexibility also allows individual utility functions to deviate from the fitness functions of genetic replicators. Hence, it can be difficult to determine the extent to which a behavior has been shaped by evolving populations of genes, experience, or both.

Since *neural replicators* exist within embodied systems, embedded within external environments, neural evolution exhibits the reciprocal functional relations characteristic of the interplay between genetic natural selection and niche construction. However, the far greater rate of neural evolution amplifies the inherent non-linearity of systems governed by reciprocal causality. In this way, utility maximizing dynamics may be more accurately described with *evolving utility generative models* that update themselves on the basis of experience. These theoretical generative models would specify *probabilistic utility landscapes*, defining the expected utilities of various action-outcome associations, as well as the adaptive significances for underlying patterns (i.e., neural replicators). That is, neural patterns are more likely to replicate their forms if they contribute to outcomes with higher expected utilities. These utility landscapes are probabilistic because expectations are based on inferences—both explicit and implicit—that influence which patterns are likely to predominate within the nervous system, thereby shaping action selection. Since different actions contribute to different experiences, which shape further action-influencing inferences, this process of reciprocal causation allows for self-directed evolutionary dynamics.

The process of estimating utility generative models is as complicated and difficult as it sounds, and in more rarified forms this ability seems to be a uniquely human capacity known as "theory of mind" (Call & Tomasello, 2008; Kaminski, Call, & Tomasello, 2008; K. Markram & Markram, 2010). In order to truly understand an agent's actions, it is necessary to understand its system of beliefs and desires, which are latent variables to which we have relatively incomplete access. Further, the limited complexity and precision with which we are able to model these variables further restricts our ability to predict behavior. Moreover, although introspection provides us with a rich assortment of memories, fantasies, and feelings, we remain mysterious even to ourselves in countless respects. Indeed, people are notoriously inaccurate in predicting their future desires via "affective forecasting" (Gilbert, 2006). And these complications become even more challenging in the context of social interactions (Faratin, Klein, Sayama, & Bar-Yam, 2002).[5]

In describing how genetic evolution produced organisms with nervous systems that support further evolutionary processes, MEDO provides an overarching framework for understanding the development of goal-oriented behavior. Selective processes are analyzed on all levels capable of influencing outcomes, the unique replicative dynamics within each level, as well as interactions between levels. However, in order to predict the relative likelihoods of particular outcomes emerging from these dynamics, it is necessary to be precise in specifying the sources of utility that drive these systems.

## GENERALIZED DARWINISM AND EDT: SELECTIVE PRESSURES AS PREFERENCES; PREFERENCES AS SELECTIVE PRESSURES

*"Suppose nothing else were 'given' as real except our world of desires and passions, and we could not get down, or up, to any other 'reality' besides the reality of our drives - for thinking is merely a relations of these drives to each other; is it not permitted to make the experiment and to ask the question whether this 'given' would not be sufficient for also understanding on the basis of this kind of thing the so-called mechanistic (or 'material') world?"*
        –Friedrich Nietzsche (1844-1900)

In light of the phenomena we have considered, MEDO identifies the following correspondences when characterizing evolutionary systems in terms of generalized Darwinism and the minds of individual agents in terms of EDT:

1. The expected changes in overall fitness from particular evolutionary outcomes (i.e., adaptive significances) correspond to expected changes in overall value from particular action-outcome associations (i.e., expected utilities). That is, *adaptive significances in evolutionary systems are comparable to expected utilities in the minds of individual agents*.
2. Differential adaptive significances (i.e., selective pressures) correspond to differential expected utilities (i.e., preferences). That is, *selective pressures in evolutionary systems are comparable to preferences in the minds of individual agents*, the relative strengths of which influence likelihoods for various outcomes.
3. The strengths of relative selective pressures (i.e., fitness functions) correspond to the strengths of relative valuations (i.e., utility functions). That is, *fitness functions characterizing differential selection within evolutionary systems are comparable to utility functions characterizing differential valuation within the minds of individual agents.*
4. Fitness functions estimated from observed adaptations and their relative frequencies (i.e., inferred selective pressures) correspond to utility functions estimated from observed choices (i.e., revealed preferences). That is, *the process of inferring selective pressures from evolutionary outcomes is comparable to the process of inferring preferences from individual choices*.

---

[5] As will be described later, social *intelligence* may have been the primary selective pressure for the increased nervous system complexity that would eventually give rise to culture. We will also describe how cultural evolution enhances the power of optimizing neural systems, thereby enabling further dynamics that must be included in any evolutionary analysis of human psychology.

5. The magnitudes of positive and negative adaptive significances (i.e., average increases and decreases in overall fitness, respectively) correspond to the magnitudes of positive and negative expected utilities (i.e., reward and punishment expectations, respectively). That is, *average fitness changes in evolutionary systems are comparable to expected utility changes for individual agents*, both of which can potentially shape future selection in circularly causal feedback processes.

6. Graphical representations of fitness function values with various feature permutations (i.e., fitness landscapes) correspond to graphical representations of utility function values with various outcome permutations (i.e., utility landscapes). That is, *fitness landscapes defining adaptive significances within evolutionary systems are comparable to utility landscapes defining expected utilities within the minds of individual agents.*

7. Peaks in fitness landscapes (i.e., evolutionary outcomes with maximum adaptive significances) correspond to peaks in utility landscapes (i.e., action-outcome associations with maximum expected utilities). That is, *selecting evolutionary outcomes that maximize fitness is comparable to choosing actions where outcomes are expected to maximize utility.*

8. The speed and reliability with which evolutionary optimization processes discover fitness maxima (i.e., algorithmic efficiency) correspond to the speed and reliability with which minds discover utility maxima (i.e., effective intelligence). That is, the *relative efficiencies of various evolutionary algorithms are comparable to the relative effective intelligences of various minds*.

9. Failures of evolutionary systems to maximize fitness in shaping outcomes (i.e., non-intentionality and blind optimization) correspond to failures of agents to maximize utility through choices (i.e., unconscious dynamics and bounded rationality). That is, *non-intentionality and blind optimization in evolutionary systems are comparable to unconscious dynamics and bounded rationality in the minds of individual agents*.

In analyzing the neural bases of goal-oriented behavior, we will demonstrate that it is more than a fortuitous coincidence that so many parallels can be found between generalized Darwinism and EDT. Rather than mere happenstance, these correspondences exist because minds are evolutionary systems and preferences function as selective pressures with respect to the neural patterns they influence. Further, since evolution's ability to produce *design without a designer* can be conceptualized as *optimization without an intentional optimizer*, the process of natural selection can similarly be conceptualized as *choice without a chooser*. Moreover, these pseudo-choices further imply that evolutionary systems allow for *reward without an experiencer*, and even *expectancy without a reasoner*.

## RESOLVING THE FREE WILL PROBLEM USING MEDO: EMERGENCE, EVOLUTIONARY EXPLANATIONS, AND CAUSATION ON ULTIMATE AND PROXIMATE LEVELS

*"Autonomous man serves to explain only the things we are not yet able to explain in other ways. His existence depends upon our ignorance, and he naturally loses status as we come to know more about behavior. The task of a scientific analysis is to explain how the behavior of a*

*person as a physical system is related to the conditions under which the human species evolved and the conditions under which the individual lives."*
       –B.F. Skinner (1904-1990)

*"'Teleology is like a mistress to a biologist: he cannot live without her but he's unwilling to be seen with her in public.' Today the mistress has become a lawfully wedded wife. Biologists no longer feel obligated to apologize for their use of teleological language; they flaunt it. The only concession which they make to its disreputable past is to rename it 'teleonomy'."*
       –David Hull (1935-2010)

Although EDT and generalized Darwinism share many correspondences, they also exhibit a fundamental difference with respect to intentionality. More specifically, agent choices can be governed by conscious purposes (i.e., *teleology*), while natural selection can only produce illusory goal-directedness (i.e., *teleonomy*). In this way, all minds are evolutionary systems, but not all evolutionary systems are minds. However, replicative dynamics drive both mental and evolutionary systems. If mental states result from the functioning of neural patterns, then it could be argued that goal-directed causation is always illusory, whether applied to selective pressures or individual preferences.

MEDO rejects this sort of analysis, as it fails to model phenomena where reciprocal causation gives rise to *emergent properties* with new causal significances (Dennett, 2003; Hofstadter, 1979; E. Thompson, 2006). That is, *system-wide properties are influenced by synergistic interactions among constituents, but these interactions are also influenced by system-wide properties*. This functional interdependence means that wholes and parts are inseparable, since the causal significances of parts have bidirectional relationships with their systemic organization into wholes. Although functional properties of mental and evolutionary systems both emerge from replicative dynamics, these causal analyses do not refute the significances of higher-level properties.[6] Indeed, complex systems are best understood by considering multiple levels of analysis, as well as the particular correspondences between these levels.

With respect to characterizing biological phenomena, Tinbergen (1963) distinguished between four types of explanations (Tinbergen, 1963): 1) *mechanism* explains how particular forms result in specific functional properties; 2) *adaptation* explains how specific functional properties (i.e., mechanisms) influence fitness; 3) *ontogeny* explains how developmental processes result in systems having particular forms with specific causal properties (i.e., adaptations); 4) and *phylogeny* explains how specific selective pressures changed the frequencies of particular developmental processes (i.e., ontogenies).

These four types of explanations can be clearly identified in our previous thought experiment with evolving populations of fruit-eating organisms: 1) *mechanism* corresponds to the physiological and behavioral characteristics related to obtaining nutrition from various fruits (e.g. strong jaw bones for chewing apples, or strong finger bones for peeling oranges); 2) *adaptation* corresponds to viewing these fruit-related specializations in terms of their fitness consequences (e.g. obtaining nutrition for survival and reproduction); 3) *ontogeny* corresponds to the developmental processes that gave rise to these specializations (e.g. relative strengthening of finger or jaw bones through mineral deposition); 4) and *phylogeny* corresponds to the history of selection through which particular developmental and mechanistic properties either increased or decreased in prevalence (e.g. differential replication of organisms with strong fingers and relatively weak jaws).

---

[6] The mind can no more (and no less) be reduced to neural patterns than a hurricane to a collection of water droplets.

Tinbergen further divided biological explanations into *ultimate causes* and *proximate causes*, both of which are valid on different levels of analysis. The proximate level is concerned with "how questions" and includes both mechanism and ontogeny (e.g. the functioning and development of specific physiological processes that allow organisms to obtain nutrition from various fruits). The ultimate level, in contrast, is concerned with "why questions" and includes both adaptations and phylogeny (e.g. oranges being more valuable nutrition sources than apples in the environment of evolutionary adaptation, leading to differential specializations). However, these two levels interact in that proximate ontogenies and resulting mechanisms are the consequence of adaptations being selected over the course of phylogeny. Further, the adaptive significance of proximate mechanism influences subsequent evolution, thus influencing ultimate level processes that shape the prevalence of future adaptations (e.g. niche construction whereby fruit eaters co-evolve with fruit-producing trees). Thus, *phylogeny is the selection of ontogenies that create mechanisms whose adaptive significance shapes future phylogeny*.

In this way, selective pressures are ultimate-level descriptions of causes that lead to proximate-level developmental processes and their resulting mechanisms (Richard Dawkins, 1996b; Galef, 2009; Geisler & Diehl, 2002). Similarly, *with respect to individual agents, preferences can be viewed as ultimate-level causes that select for the development of proximate-level choices* (e.g. choosing oranges because of their higher expected utility, relative to apples). Further, these proximate-level *choices can be viewed as ultimate-level causes that select for the proliferation of neural patterns, whose interactions create proximate level causes for the realization of particular action sequences* (e.g. the desire to eat a particular orange resulting in an intricate series of muscle movements to remove the peel).[7]

In addition to personal motivation, a further set of ultimate-level causes can be identified in the ways that natural selection shaped behavioral tendencies over evolutionary history. However, as a blind optimization process, selective pressures apply to the consequences of actions, rather than specific behaviors (Richard Dawkins, 1996a; Dennett, 1996, 2009). As a result, fitness can be maximized by a variety of proximate-level mechanisms, for which genetic involvement may be highly indirect. Indeed, although behavioral consequences may increase fitness for replicators that cause those behaviors on ultimate and proximate levels—genes and neural patterns, respectively—the subjective goals of individual organisms may have little resemblance to the adaptive significances to which they correspond (e.g. desiring oranges for pleasurable eating experiences, rather than their nutritional importance in the environment of evolutionary adaptation).

---

[7] The manner by which intentions are translated into actions will be discussed in greater detail in the section, "Cortical control systems."

In order to better understand these multi-level causes, in the following sections we will describe the evolutionary developmental origins of nervous systems, the ways that they are impacted by genetic and cultural replicators, and the similarities and differences between these various selective processes. Then after discussing the mechanisms by which neural evolution produces intentional actions, we will be able to more precisely evaluate the conditions that enable autonomous behavior.

## MULTILEVEL SELECTION: EVOLUTION, DEVELOPMENT, LEARNING

### EVOLUTIONARY DEVELOPMENTAL BIOLOGY

#### UNITS OF SELECTION: GENE-CENTERED AND ORGANISMIC VIEWS OF EVOLUTION

*"Fitness is a property, not of an individual, but of a class of individuals — for example homozygous for allele A at a particular locus. Thus the phrase 'expected number of offspring' means the average number, not the number produced by some one individual. If the first human infant with a gene for levitation were struck by lightning in its pram, this would not prove the new genotype to have low fitness, but only that the particular child was unlucky."*
    –Maynard Smith (1920-2004)

*"The essence of the genetical theory of natural selection is a statistical bias in the relative rates of survival of alternatives (genes, individuals, etc.). The effectiveness of such bias in producing adaptation is contingent on the maintenance of certain quantitative relationships among the operative factors. One necessary condition is that the selected entity must have a high degree of permanence and a low rate of endogenous change, relative to the degree of bias (differences in selection coefficients). Permanence implies reproduction with a potential for geometric increase."*
    –George Williams (1926-2010)

As with all other organs, nervous systems consist of interacting tissues shaped by evolution according to two general categories of selective pressures: 1) maintaining and expanding the *homeostatic boundaries* of organisms, or 2) facilitating *reproduction* (A. Damasio, 2000, 2003; Panksepp, 1998). Homeostatic boundaries are the range of conditions under which organisms can function and maintain essential processes for self-preservation (Heylighen, 1992; Heylighen & Joslyn, 2003; Rudrauf, Lutz, Cosmelli, Lachaux, & Le Van Quyen, 2003; Waddington, 1953a). Expanding these boundaries enables increased flexibility in the Darwinian struggle for existence wherein individuals compete for access to limited resources (e.g. food, territory) within their respective niches (Aldrich et al., 2008; Gause, 1934). However, this competitive advantage only translates into evolutionary fitness if it also contributes to reproductive success for organisms with genes supporting these adaptations.

For example, different organisms have varying nutritional requirements that must be satisfied if they are to survive. These requirements represent homeostatic boundaries in that unless adequate nutrition is obtained, the organism will be unable to continue functioning and will eventually die. Yet, these boundaries include more than just the current physiological status of the organism, but also the availability of nutrient sources. Nervous systems can expand these boundaries if they make organisms more effective at obtaining food in a wider variety of circumstances. Nervous systems can also result in behaviors—and even technologies—that allow a wider variety of food sources to be metabolized (e.g. red colobus monkeys eating charcoal in order to neutralize toxins (Struhsaker, Cooney, & Siex, 1997)). However, biological features that enable this flexibility will only have positive adaptive significance if they also result in increased replication of the underlying genetic information.

From the *gene-centered view* of evolution, individuals are mere "vehicles" for the "true replicators" instantiated in genes, which in turn code for the phenotypic properties that define particular organisms (Richard Dawkins, 1976; Williams, 1966). That is, genes influence the development of organ systems in ways that generally increased the likelihood of spreading those genes within the environment of evolutionary adaptation. In this way, genes are able to increase the likelihood that they will be replicated by shaping phenotypes such that organisms are more likely to survive and reproduce. Indeed, evolutionary fitness corresponds to the degree to which adaptations maximize the overall replication of specific genetic patterns (i.e., *inclusive fitness*), regardless of whether or not it promotes the survival and reproduction of the individual organisms containing those genes (Richard Dawkins, 1999; Nowak, 2006).

Evolutionary theorists have intensely debated this gene-centered view, with alternative proposals suggesting that the *sum-total of phenotypic traits* constituting the organism is the most appropriate *unit of selection*, or that selective pressures must be considered on *multiple levels of organization* (Gould, 1992; Maynard Smith & Szathmáry, 1995). From an *organismic perspective*, both genotypes and phenotypes exist within the context of (reciprocally) causal networks where individual features are meaningless outside of their interactions within *functional cycles,* wherein various systems synergistically interact (Huang, 2012; Lane, 2005; Slack, 2002; Von Uexküll, 1957). Further, genes are only capable of replicating via selection upon phenotypes, which emerge through complexly determined developmental processes (i.e., ontogeny) (Northcutt, 1990; Wagner & Laubichler, 2004).

Genes figure centrally in development, but these processes are also impacted by non-genetic factors such as environmental circumstances. Indeed, genes are only able to impact phenotypes by influencing dynamics within and between cells that function in the *hierarchical contexts* of tissues, organs, organ systems, organisms, and ecosystems (Barbieri, 2008; Jablonka & Lamb, 2007; Kohl, 2012). Further, the functioning of *gene networks* within and between cells is a product of their unique developmental histories within the context of the overall organism, which influence partially heritable *epigenetic* processes that determine the levels of expressions for various genes (Haslberger, Varga, & Karlic, 2006; P. A. Jones & Takai, 2001; E. K. Murray, Hien, de Vries, & Forger, 2009). In this way, genotypes do not deterministically specify phenotypes, but instead *probabilistically* *influence* which phenotypes are likely to emerge via interactions within and between multiple hierarchical levels of organization.

More specifically, the organismic property of *phenotypic plasticity* describes how a given genotype can produce a different phenotypes in response to varying environmental conditions (Badyaev, 2009; Baldwin, 1896; Ernande & Dieckmann, 2004; Johnston & Gottlieb, 1990; Lessells, 2008; Price, Qvarnström, & Irwin, 2003; Waddington, 1953a, 1953b). This phenotypic variability provides a diversity of forms that can influence fitness within a niche. The "Baldwin effect" refers to the phenomenon of this plasticity-based differential fitness resulting in selection for the heritable aspects of the phenotypic varieties that are most highly adaptive. Alternatively, this *genetic accommodation* of traits in response to novel selective pressures can also be conceived of in terms of *genetic assimilation*, whereby the developmental process is *canalized* (i.e., shaped) in a way that it is more likely to produce phenotypes that were adaptive in the environment in which they were selected.[8] Canalization is a complimentary principle to phenotypic plasticity, in that it refers to the ability of similar phenotypes to result from a variety of environments and genotypic configurations. While phenotypic plasticity supports adaptive flexibility, canalization supports the robustness of specific functions. These concepts clearly demonstrate the probabilistic relationship between genotypes and phenotypes, providing examples of emergent properties that must be considered in assessing the functional significance of genetic replicators.

However, the organismic and gene-centered perspectives are compatible when viewed as complimentary analyses on different levels of abstraction. For example, in the theory of *extended phenotypes*—beyond influencing the physiological processes of individual organisms containing those genes—phenotypes have blurry boundaries in that they include a variety of knock-on effects, such as the impacts of organisms on their environments (e.g. beaver dams), which can also include influences on the behaviors of other organisms (e.g. toxoplasmosis parasitism) (Richard Dawkins, 1999; Dennett, 2009; House, Vyas, & Sapolsky, 2011). Although genetic maximization is the primary determinant of evolutionary outcomes in these models, they also encompass the complex interactions that mediate the relationship between genotypes and selective pressures.

Interactive dynamics notwithstanding, the *temporal stability* of genetic replicators is a *necessary condition for optimized complexity to accrue* via natural selection. Since genes— and to varying degrees, heritable epigenetic patterns—are the only aspects of organisms that persist over multiple generations, the relative proliferation of these replicators are the "bottom-line" for *cumulative selection* (Dennett, 1996). In this way, differential genetic replication—broadly construed to include epigenetic inheritance—is the ultimate determinant of adaptive significances that allow populations to climb fitness landscapes on evolutionary timescales (i.e., phylogeny) (Richard Dawkins, 1996a, 1997). However, this differential replication is determined by the functional properties of phenotypes mediating the relationship between selective pressures and genotypes. In this way, both the gene-centered and organismic perspectives are necessary in order to fully characterize evolutionary dynamics.

---

[8] Although distinctions can be made between the Baldwin effect and genetic assimilation (Crispo, 2007), for our purposes, the latter can be considered to be a restatement of the former from an evolutionary developmental perspective (King & Bjorklund, 2010; Mills & Watson, 2005; Pigliucci, 2009; Wagner & Laubichler, 2004).

## RESOLVING THE UNIT OF SELECTION DEBATE USING MEDO: SELFISH REPLICATORS AND THEIR GAME-THEORETIC INTERACTIONS WITHIN EVOLUTIONARY SYSTEMS

*"We have the power to defy the selfish genes of our birth and, if necessary, the selfish memes of our indoctrination. We can even discuss ways of deliberately cultivating and nurturing pure, disinterested altruism - something that has no place in nature, something that has never existed before in the whole history of the world. We are built as gene machines and cultured as meme machines, but we have the power to turn against our creators. We, alone on earth, can rebel against the tyranny of the selfish replicators."*
        –Richard Dawkins (1941-present)

*"It is not from the benevolence of the butcher, the brewer, or the baker that we expect our dinner, but from their regard to their own interest."*
        –Adam Smith (1723-1790)

Whether considered in terms of genotypes or phenotypes, evolutionary fitness is determined by the reproductive success of all organisms sharing those traits. Yet, *sexual reproduction* allows different versions of genes (i.e., alleles) to be combined in different ways in different organisms (Ridley, 1993). Since each allele can be thought of as a separate heritable element, the adaptive significances of traits are partially independent from other traits, as well as the reproductive success of any given individual. In light of this partial independence with respect to selective pressures for different genes, Dawkins (1976) argued for using a "gene's-eye view" that conceptualizes evolution in terms of the activity of "selfish" replicators (Richard Dawkins, 1976). These replicators are considered to be selfish in that they function as if they possessed a singular goal for maximizing their own proliferation, with all other outcomes being means to this end.

This sort of goal-directedness is clearly illusory with respect to genes, but all teleonomy is described with EDT in the MEDO framework. If a utility function were ascribed to a replicator, it would necessarily be self-interested by virtue of the characteristics that define it as a replicating phenomenon. That is, replicators behave as if they value the replication of their forms. Yet, it is frequently more rational for selfish actors to coordinate in avoiding competition and cooperatively maximizing utility, depending on *game-theoretic* considerations that *define incentive structures for interactions* (Faratin et al., 2002; Von Neumann & Morgenstern, 1944). Biological replicators must coordinate their functioning if they are to ensure the survival of the organismic "vehicles" upon which they depend for their proliferation, and cooperative synergy may be just as fundamental as competition in shaping evolutionary outcomes (Nowak, 2006).

However, MEDO uses EDT to characterize all teleonomic phenomena in terms of natural selection, rather than the replicative properties of individual patterns. Although replicators can be thought of as *selecting* interactions in which their forms are perpetuated, they are also *selected by* emergent causal dynamics of systems that enable replication. That is, broader evolutionary systems provide contexts that determine the relative success of various replicative dynamics. These systemic properties could be thought of as defining incentive structures for the interactions of selfish replicators, with fitness maxima representing game-theoretic equilibria (Maynard Smith, 1982; Nowak & Sigmund, 2004; J. M. Smith, 1976).[9] In addition to the *replicator's-eye view*, MEDO adopts the *evolutionary-system's-eye view*, focusing on the emergent dynamics that determine selective pressures. This alternative focus does not refute the validity of the gene-centered view, but subsumes it—as well as evolutionary game theory—thereby providing a compatible level of analysis for understanding these complex phenomena.

By applying EDT to evolutionary systems, MEDO clarifies the relationships between gene-centered and organismic perspectives. To the extent that some phenotypic traits increase fitness more than others, the relative strengths of the selective pressures influencing differential replication correspond to an abstract utility function—or fitness function—describing the relative likelihoods that the underlying genes (and heritable epigenetic patterns) will either increase or decrease in prevalence. From the organismic perspective, this evolutionary utility function would be expressed in terms of the fitness consequences of the phenotypes, which mediate the relationships between selective pressures and genotypes. However, the adaptive significances of phenotypic traits are completely conditional upon their impacts on differential genetic proliferation. These forms will not be replicated unless their underlying genetic information is passed on to future generations.

The utility function of an economic agent is meaningless outside of its ability to make choices. Similarly, genetic utility must be considered in terms of the resulting phenotypes, which influence likelihoods that various genes will be replicated. Hence, phenotypes can be viewed as revealed preferences of evolutionary systems in which expected utilities are defined by the likelihoods that the underlying genes will either increase or decrease in prevalence. Differential genetic replication shapes phenotypes, but the emergent properties of phenotypic traits also shape gene frequencies on the population level. Indeed, reciprocal causation defines the relationship between genotypes and phenotypes in natural selection. Thus, MEDO clearly demonstrates that the gene-centered and organismic perspectives are not just compatible, but complimentary.

MEDO considers multiple units of selection to varying degrees, depending on context. A Darwinian/EDT analysis is applied to all levels at which significant replicative dynamics can be identified, as this will determine the extent to which systems can be characterized as having evolutionary properties. Indeed, there are fuzzy boundaries between replicative and non-replicative phenomena, in that patterns exist in terms of their ability to impact particular causal systems (Jablonka & Lamb, 2007; Shay, 2010), and these impacts are always a matter of degree, even when they contribute to emergent differences in kind. All of the defining characteristics of natural selection (i.e., replication, variation, and selection) can only be quantified with limited precision, and to the extent that this quantification is possible, functional significance can be difficult to determine. For example, although mutation rates can be compared, and population-level changes in allele frequencies estimated, the strength of selection pressures on different phenotypes changes continuously on multiple levels. Or with respect to cultural evolution, selective pressures may be highly variable, and precise replicators may be unidentifiable, but natural selection still occurs on the level of culture (Aldrich et al., 2008; Nowak, 2006; Nowak & Sigmund, 2004; N. S. Thompson, 1998; Traulsen & Nowak, 2006; Wilson, 1994). Something may be difficult to precisely quantify, yet still have significant causal properties.

---

[9] More specifically, evolutionary outcomes would tend to represent mixed-strategy Nash equilibria, where all "players" (i.e., replicators) select their best responses, given the likely best responses of all other players.

In light of these epistemological challenges, we may be limited in the extent to which we can specify the relative contributions of different selective dynamics to shaping motivation. However, these difficulties can be partially ameliorated by using evolutionary and developmental constraints to limit the range of plausible hypotheses, thereby enhancing inferential power. In the following discussions, by considering the means by which selective pressures interact with developing nervous systems, we will attempt to estimate the relative extent to which various dynamics are likely to influence preferences.

## EVOLUTIONARY-DEVELOPMENTAL UTILITY FUNCTIONS: PHYLOGENY AS ONTOGENY; ONTOGENY AS PHYLOGENY

*"My idea is that every specific body strives to become master over all space and to extend its force (its will to power) and to thrust back all that resists its extension. But it continually encounters similar efforts on the part of other bodies and ends by coming to an arrangement ('union') with those of them that are sufficiently related to it: thus they then conspire together for power. And the process goes on."*
     –Friedrich Nietzsche (1844-1900)

*"Thus, from the war of nature, from famine and death, the most exalted object which we are capable of conceiving, namely, the production of the higher animals, directly follows. There is grandeur in this view of life, with its several powers, having been originally breathed into a few forms or into one; and that, whilst this planet has gone cycling on according to the fixed law of gravity, from so simple a beginning endless forms most beautiful and most wonderful have been, and are being, evolved."*
     –Charles Darwin (1809-1882)

Both the temporal stability of genetic replicators and their phenotypic consequences are necessary conditions for the accumulation of optimized complexity through phylogeny. However, the ontogeny (i.e., development) of individual organisms represents an *additional selective process* that always mediates the production of phenotypes from genotypes. More specifically, since the maturation of organisms involves replication, variation, and selection of cells and other biological patterns—in addition to being a product of phylogeny—*development is necessarily an evolutionary process* in its own right, sui generis. Hence, the developmental process selects among interacting *ontogenetic replicators*, thereby resulting in particular mechanisms, the sum-total of which constitute particular phenotypes.[10]

As with all other evolutionary systems, MEDO models development as a kind of natural selection, thereby constituting an optimization process that can be characterized in the language of EDT. To the extent that particular gene networks influence the development of particular phenotypes, the relative strengths of organizing processes correspond to an abstract utility function—or fitness function—describing the relative likelihoods for the emergence of different phenotypes. Hence, MEDO formalizes Waddington's (1953) proposal that the relative likelihoods of development being shaped in different ways correspond to the topology of *epigenetic landscapes* (Huang, 2012; Waddington, 1953a, 1953b). That is, if gene networks specify the kinds of "utility" that influence developmental processes, then the theoretical graphs of these *ontogenetic utility functions* can be visualized as high-dimensional landscapes in which biological patterns evolve to create organisms with particular combinations of phenotypic traits. Thus, the evolution of genetic replicators through differential reproduction (i.e., phylogeny), of phenotypes through development (i.e., ontogeny), and of agents through experience (i.e., goal-oriented behavior) can all be represented by their respective theoretical utility functions and landscapes, corresponding to the expected prevalence of different outcomes.

*Ontogenetic utility* can be viewed as an additional ultimate-level cause for the creation of phenotypes, with particular cellular and systems-level physiological processes acting as proximate-level means of realizing the implicit preferences of gene networks. Hence, genes can act as ultimate-level biological causes in two senses: 1) *phylogenetically* acting as replicators that shape phenotypes to maximize their proliferation; 2) *ontogenetically* acting as causal networks in order to shape developmental processes that maximize the replication of particular phenotypic patterns. Ontogeny plays a dual evolutionary role as a proximate-level implementational mechanism for genetic replicators, and as an ultimate-level cause in-and-of-itself. In this way, phylogeny can be understood as a kind of *meta-evolution* selecting among *genotype-phenotype developmental pathways*, which themselves function as evolutionary systems selecting among ontogenetic replicators.

Since the particular configurations of gene networks influencing ontogeny result from a history of phylogeny, phylogenetic and ontogenetic utility functions (i.e., abstract descriptions of what is likely to be adaptive and what is likely to develop, respectively) will tend to be highly similar under normal circumstances. That is, organisms will tend to develop such that they function in ways that increase replication of the genes contributing to the formation of their particular phenotypes. However, these utility functions are never perfectly aligned for a variety of reasons. For example, while environmental conditions may change rapidly and non-linearly, genotype compositions tend to change gradually— although not always—through cumulative selection (Eldredge & Gould, 1972; Gilad,

---

[10] Importantly, since phenotypes have blurry boundaries, ontogeny extends beyond the physiological maturation of particular organisms—processes that continue over entire lifespans—including factors such as the formation of behavioral interactions and other niche modifications with adaptive significances.

Oshlack, Smyth, Speed, & White, 2006). If a population contains adaptive genotype-phenotype pathways, then these multilevel evolutionary utility functions are likely to become increasingly aligned as these replicators spread through the population over multiple generations.

Alternatively, if no adaptive genotype-phenotype pathways are present, then time will be needed for the requisite mutations to arise. Most mutations are neutral in their effects, but when they do have functional significance, they are most likely to be (mildly) deleterious, and only rarely infer an adaptive benefit (Fay, Wyckoff, & Wu, 2001). More specifically, a single genetic alteration can change the functional interactions of gene networks in multiple ways (i.e., epistasis), thus potentially creating nonlinear changes in phenotypes. If such a nonlinear change occurs, it is prima facie unlikely to be adaptive in light of the fact that physiology has been subject to a distributed optimization process, operating over geological timescales.[11] However, if an adaptive phenotype-inducing mutation does arise, then differential rates of reproduction for organisms containing those mutations is likely to align phylogenetic and ontogenetic utility functions.

Indeed, the *degree of evolvability of particular mechanisms* is determined by their abilities to minimize discrepancies between selective pressures and developmental biases. In order to ascertain the likelihood of a mechanism evolving, all relevant phylogenetic and ontogenetic dynamics must be considered simultaneously, on all levels at which functional interactions take place. Although we are necessarily limited in the rigor with which such an analysis can be conducted, relative evolvability can be qualitatively estimated. The number and strength of selective pressures for a particular adaptation influences the probability that one or more proximate mechanisms are likely to develop. The number and variety of developmental pathways that could produce a proximate mechanism influence the probability that one or more of those pathways will be selected. In this way, as we will demonstrate in the following sections, the mutual constraints of phylogeny and ontogeny influence the degree to which we can predict the relative likelihoods for evolution giving rise to different mechanisms.

## THE EVOLUTIONARY ORIGINS OF HUMAN MINDS

### EVOLVING NERVOUS SYSTEMS

*"Let us assume that the persistence or repetition of a reverberatory activity (or "trace") tends to induce lasting cellular changes that add to its stability…. When an axon of cell A is near enough to excite a cell B and repeatedly or persistently takes part in firing it, some growth process or metabolic change takes place in one or both cells such that A's efficiency, as one of the cells firing B, is increased."*
    –Donald Hebb (1904-1985)

---

[11] Consequently, molecular error-correction mechanisms exist to ensure that mutations are relatively rare events, even though there can be adaptive variability in mutation rates (Lynch, 2010, 2011), and macro-mutations occasionally take place (Larsson et al., 2008).

*"If the inputs to a system cause the same pattern of activity to occur repeatedly, the set of active elements constituting that pattern will become increasingly strongly interassociated. That is, each element will tend to turn on every other element and (with negative weights) to turn off the elements that do not form part of the pattern. To put it another way, the pattern as a whole will become 'auto-associated'. We may call a learned (auto-associated) pattern an engram."*
      –Gordon Allport (1887-1967)

MEDO not only analyzes development in terms of the ways that it has been shaped by evolution, but it considers neurodevelopment to represent an additional kind of evolutionary process for increasing phenotypic plasticity. Networks of neurons allow nervous tissue to have a uniquely plastic internal structure, whereby specialized cells are capable of changing their functional interconnectivity on the basis of patterns of activity that spread through these connections. These connections define "cell assemblies" or systems of neurons with functional significance for other systems (Hayek, 1952; Hebb, 1949). The activity dependent creation and re-formation of cell assemblies has been most simply expressed in the maxim: "fire together, wire together" (Shatz, 1996). As organisms interact with the world, these different assemblies form interconnections on the basis of correlated activity, and thus create auto-associated networks of increasing complexity and specificity (i.e., "engrams") (Allport, 1985; James, 1890; Schott, 2011).

Similarly to how phenotypic plasticity facilitates adaptation on the population level, neuroplasticity is a particularly powerful variety of phenotypic plasticity that facilitates adaptive behaving on the level of both populations and individuals. Under the MEDO framework, learned behaviors are considered to be "adaptive" in an evolutionary sense because the learning process itself is considered to be a form of natural selection. More specifically, neuroplasticity creates an evolutionary optimization process whereby specific patterns are selected within the nervous systems of individual organisms based on their relative abilities to facilitate dynamics that tend to maximize evolutionary fitness on multiple levels. By enabling natural selection to take place within nervous systems, evolution gave rise to selective dynamics that were able to operate far more rapidly than trial-and-error on the level of populations, and thus more effectively optimize behavior. In the parlance of EDT, to the extent that learning processes favor some patterns over others, an implicit utility function specifies the relative likelihoods that various patterns will be learned, and differential prevalence can be expected to reflect the relative strengths of these learning biases. To the extent that learned patterns correspond to behaviors, these utility functions are identical to those for agent preferences, and can be thought of as the emergent property of the sum-total of selective pressures within the nervous system.

As previously described, the relationship between an organism and its environment is always interactive, involving both continuous exchanges of matter and energy, as well as multiple levels of mutually constraining feedback processes. However, the evolution of rapid electrochemical signaling in neuronal networks allowed organisms to be even more responsive to their environments. Although the functioning of biological systems always involves circular causation (Haslberger et al., 2006; Wagner & Laubichler, 2004), nervous systems increase the non-linearity of these processes through experience-based self-modification, which in turn influences the experiences that are likely to be encountered by those modifiable systems. This capacity for self-organization enables behavioral flexibility, and it is the adaptive payoff that allowed for the evolution of adaptations as complex and metabolically intensive as nervous systems (Attwell & Laughlin, 2001; Heylighen, 1992, 1999; Heylighen & Joslyn, 2003; Howarth, Gleeson, & Attwell, 2012).

The means by which these processes of self-organization give rise to optimized functioning were first proposed by Hayek (1952), and were later described by Edelman (1987) in his theory of "Neural Darwinism" (Edelman, 1993; Hayek, 1952; McDowell, 2010; Seth & Baars, 2005; Wyckoff, 1987). Edelman's model of "neuronal group selection" proposes that neural patterns evolve, compete, and cooperate on the level of systems of connected neurons that he refers to as "ensembles" (Edelman, 1993; Wyckoff, 1987).[12] In this theory, phenotypic plasticity plays a central role in that genes do not precisely specify the anatomical features of individual neurons. Rather, genes specify developmental processes whereby dynamically altering populations of cells provide an initial diversity of forms, which can be selected from on the basis of patterns of connectivity between neuronal processes. In this way, like any other organ, genes are capable of generating highly complex internal microstructure, but with no particular cell being special a priori to the developmental process.[13] Out of this initial morphological diversity, a second evolutionary process arises whereby experience-dependent activity selects between groups of neurons on their ability to form coherent cell assemblies that are capable of mutually stimulating each other through reentrant signaling.

This Darwinian perspective helps to explain the adaptive significance of the massive neuronal pruning that takes place over the course of development, and can extend for almost three decades in humans (Gogtay et al., 2004; Sisk & Foster, 2004). Initially, an over-abundance of neurons are generated, but only those neurons that receive adequate stimulation from other neurons undergo survival-enhancing metabolic changes, such as the release of neurotrophic growth factors that inhibit the default processes of programmed cellular death (i.e., apoptosis) (T. Elliott, Maddison, & Shadbolt, 2001; Galuske, Kim, Castrén, & Singer, 2000; Galuske, Kim, Castren, Thoenen, & Singer, 1996; Hennigan, O'Callaghan, & Kelly, 2007). More specifically, groups of neurons receive this survival-enhancing stimulation when they contribute to patterns of activity that facilitate dynamics that stimulate the ensembles in which they participate. Alternatively, neurons that fail to get taken up into coherently active groups do not receive this stimulation, and thus are eliminated via apoptosis. Further, processes of synaptic modification allow individual ensembles and systems of ensembles to plastically adapt in order to more effectively compete and cooperate in these selective processes (Caporale & Dan, 2008; Song, Miller, & Abbott, 2000; van Rossum, Bi, & Turrigiano, 2000). Moreover, ongoing synaptic modification—and potentially other mechanisms such as ongoing myelin production (Ullén, 2009)—allows neuroplasticity-based evolution to continue after the rate of neuronal pruning begins to slow post-adolescence (Honig & Rosenberg, 2000).

---

[12] We will use the terms ensemble and assembly interchangeably.

[13] Indeed, if the functional significance of particular neuronal connections are unpredictable a priori—because of the sensitivity to initial conditions exhibited by complex systems (Elbert et al., 1994; Korn & Faure, 2003; Rabinovich & Abarbanel, 1998; Wolfram, 2002)—then one could argue that even in principle, genes necessarily had to rely on novel optimization processes in order to produce adaptive configurations.

Thus, for neural systems there are numerous mechanisms by which replication, variation, and selection can implement multilevel evolutionary processes. However, in order to predict what sorts of behavioral outcomes might result from these processes, it is necessary to identify the operative selective pressures as precisely as possible. Just as organ systems must operate cooperatively if the organism is to survive (Von Uexküll, 1957), neural patterns must also form functional cycles for coherent information processing to occur. In this way, the informational ecosystem of an individual mind represents the primary niche for neural replicators. In order to survive, neural patterns must be perpetuated within systems on multiple interacting levels, the sum-total of which must allow the organism to effectively interact with its environment. Thus, neural patterns obtain functional significance in the context of embodied minds, embedded within environments.

Neuronal activity corresponds to patterns of actions and perceived stimuli—both internal and external to the body—whose "isomorphic" relationships are due to connections between the nervous system and the sensors and effectors of the organism (Bridgeman & Tseng, 2011; Hayek, 1952; Hofstadter, 1979). In this way, neuronal patterns represent different aspects of the world, with the system-defined meanings of this activity ultimately grounded in the perceptions and actions of an organism coupling with its environment (Barsalou, 2008; JämsÄ, 2001; Rudrauf et al., 2003; Stella & Kleisner, 2010; Von Uexküll, 1957; Ziemke & Sharkey, 2001). However, specific connections influence the probability of the patterns that flow through them and plastically shape them, both in terms of network dynamics, as well as through actions taken by the organism resulting in specific perceptions. This action-driven perception not only results from directed sensing, but also by way of more indirect pathways of actions initiating chains of events in the world on multiple timescales.

## EVOLUTIONARY UTILITY FUNCTIONS ON THE LEVELS OF POPULATIONS AND INDIVIDUALS

To the extent that neurodevelopment can be canalized such that emergent evolutionary learning processes are more likely to favor genetically adaptive behaviors—within the respective environments of evolutionary adaptation—genetic accommodation can be expected to occur over time. On the population level, the relative likelihood of genetic assimilation facilitating these plastic behavioral phenotypes will tend to be a function of the relative degree of fitness provided, as well as the relative evolvability of these learning biases. In evolutionary psychology, the innate biasing of learning processes is referred to as "preparedness" and constitutes a means by which genetic natural selection was able to increase the probability that the utility functions of individual organisms will tend to correspond with the utility functions of genetic replicators (R. J. McNally, 1987, 1995; Mineka & Ohman, 2002; Ohman & Mineka, 2001; Seligman, 1971).

This is yet another sense in which learned behaviors can be "adaptive," except this sort of adaptive behavior corresponds to the implicit "preferences" of natural selection from the gene's-eye view. Importantly, the abstract utility functions characterizing these choices may or may not correspond to the values of agents, either in terms of their individual genetic fitness or subjective values. Additionally, when learning processes achieved sufficient levels of complexity to enable cultural evolution, even more forms of utility became relevant. As previously described, in the same way that genotype-phenotype pathways converge, and in the same way that evolution and development are mutually constraining, multi-objective optimization for learning processes necessarily involves compromise among all of the conflicting and synergistic goals that interact in systems of causal relations (Abbas, 2010; D. E. Bell, 1979; Gw et al., 1995). Thus, with respect to evolving nervous systems, all selective dynamics compete and cooperate in determining the likelihood and prevalence of outcomes, from which the inferred utility function reflects the inclusive fitness of all replicators involved, on all levels to which they can interact to significant degrees.[14]

Although the probabilistic nature of development may be a necessary consequence of the fact that "epigenetic landscapes" mediate the relationship between genotype and phenotype, phenotypic plasticity can itself have selective importance in helping populations of organisms to climb "fitness landscapes" via evolutionary adaptation (Huang, 2012; Suzuki & Arita, 2007; Wright, 1932). Indeed, population-genetic defined fitness, organismic-shaping defined developmental forces, and agent-value defined motivation can all be thought of as selective processes whose nested interactions provide mutual-constraints as they are partially integrated into a single abstract utility function that describes what will tend to be optimized.

The degree to which the utility of different replicators drives optimization dynamics will depend on the timescales and contexts under consideration. Neuronal replicators and epigenetic-phenotypic pathways will dominate on timescales of individual organisms, but unless they can contribute to genetic proliferation, these selective dynamics will not influence outcomes on the level of populations of organisms changing over geological time. However, via the processes of genetic accommodation previously described, development can be canalized in a way that it is more likely to produce evolutionary dynamics that were adaptive in the environment in which they were selected.

---

[14] Neural evolution creates optimized patterns of high complexity, and as such an engineering utility analysis is appropriate for understanding the resulting behaviors. Such an analysis is not just appropriate, but it describes a fundamental capacity that has become uniquely developed in humans over the course of genetic and cultural evolution. We constantly attempt to use the specific beliefs and desires of others in order to attempt to predict their values, and attempt to use values in order to predict specific beliefs and desires.

## THE EVOLUTION OF GENE-CULTURE CO-EVOLUTION

*"As soon as the primeval soup provided conditions in which molecules could make copies of themselves, the replicators themselves took over. For more than three thousand million years, DNA has been the only replicator worth talking about in the world. But it does not necessarily hold these monopoly rights for all time. Whenever conditions arise in which a new kind of replicator can make copies of itself, the new replicators will tend to take over, and start a new kind of evolution of their own. Once this new evolution begins, it will in no necessary sense be subservient to the old. The old gene-selected evolution, by making brains, provided the `soup' in which the first memes arose. Once self-copying memes had arisen, their own, much faster, kind of evolution took off. We biologists have assimilated the idea of genetic evolution so deeply that we tend to forget that it is only one of many possible kinds of evolution."*
         –Richard Dawkins (1941-Present)

Although there are reasons to believe that a dynamic probabilistic model is particularly apt for characterizing the changing utility functions of neural evolution, it also applies to all of the previously described engineering analyses of biological evolution involving circular causation. However, this complication can be greatly simplified in light of the fact that before cultural evolution appeared, gene-pathways were the only replicators capable of accumulating appreciable degrees of complexity on evolutionary time scales. Nervous systems greatly expanded phenotypic plasticity, but without robust transmission of this information between organisms, optimized complexity was limited in its ability to accrue. Thus, for most of evolutionary history, the existence of nervous systems did not change the fact that the utility function of natural selection was completely conditional upon genetic transmission, and thus maximization of gene frequencies on the population level dominated the optimization processes that shaped biological organisms.

However, increasingly complex nervous systems allowed for increasingly complex social relationships between organisms and groups of organisms.[15] To the extent that varying features of these groups had differential effects in their ability to contribute to group preservation over time—corresponding to the definition of replication for generalized Darwinism—these groups would have represented a partially independent source of evolutionary utility. These selective processes operated on time scales where they were able to accrue optimized complexity and drive behavioral dynamics (N. S. Thompson, 1998; Wilson, 1994), and thus drive the evolution of neural replicators both between and within individual organisms. Similarly, non-genetic replicators increased reliance on navigating social alliances, thus providing a selective pressure for further intelligence (Lefebvre, 2012; L. McNally, Brown, & Jackson, 2012; Reader, Hager, & Laland, 2011).

---

[15] This should not be taken to imply that organisms with simple nervous systems are unable to engage in complex social interactions, as are clearly evidenced in the behavior of eusocial insects (Anfora et al., 2011; Patalano, Hore, Reik, & Sumner, 2012). However, the interactions between eusocial insects are more likely to be governed by processes that depend on relatively simple algorithms, compared with the rich modeling of other organisms made possible by the increased nervous system complexity of mammals. Alternatively, in the study of non-linear deterministic systems, one of the most significant—and surprising—findings to emerge is that even simple algorithms can yield enormous complexity (Wolfram, 2002).

To the extent that variations of behavioral patterns resulted in differential preservation of those patterns over time, a sort of proto-cultural evolution represented the emergence of another partially independent source of utility, in addition to the fitness of replicating groups and the genes of group members. None of these forms of utility were uniquely independent in that they mutually influenced the degree to which optimized complexity was able to accrue for each type of replicator. Indeed, not only were these replicators highly dependent upon each other, but their interactions enabled niche construction to occur with unprecedented rapidity via gene-culture coevolution and cultural group-selection (A. Bell, 2010; Henrich, 2004; Lansing & Fox, 2011; Rendell, Fogarty, & Laland, 2011).

Once these synergistic selective pressures were able to produce shared attention and intentionality, imitative learning, and symbolic cognition and communication, the necessary conditions for human-like cultural evolution had been achieved (Tomasello, 1999). Regardless of whether or not cultural natural selection is thought of in terms of replicating memes (Richard Dawkins, 1976; Dennett, 1992), this new optimization process represents a major transition in evolution (Jablonka & Lamb, 2007; Maynard Smith & Szathmáry, 1995). Not only did it radically increase the power of gene-culture coevolution and cultural group-selection, but it also allowed for new forms of natural selection to arise within individual nervous systems, and thus new types of utility (i.e., explicit meanings) were capable of shaping dynamics.

Culture allows the processes of selection that take place within individual minds to be extended to other minds as well. Cultural replicators are capable of exploring multiple permutations of problems in parallel, as well as horizontally transmitting successful solutions between individuals.[16] Additionally, robust transmission of information between individuals allows the optimized complexity of patterns within nervous systems to be extended over multiple generations. As previously discussed, temporal persistence of replicating dynamics is a necessary condition for specialized complexity to accrue. Thus, increased parallelism and multi-generational persistence vastly increased the power of culture as an optimization process.

However, although an aspect of culture can spread because it is useful to individuals, it can also spread simply because the underlying patterns are good at replicating themselves, or because of chance. To the extent that units of selection can be identified for cultural evolution, the primary niche for these replicators would be the minds of individual organisms. Just as was previously described in applying an EDT analysis to evolutionary developmental biology more generally, within the nervous systems of individual learners, multiple forms of replicators all compete and cooperate in attempting to shape overall dynamics. For agents participating in culture construction, utility dynamics within their nervous systems are the result of these selective processes operating on multiple time scales. Thus, culture is not necessarily optimized for the utility of any given individual or group of individuals, but for the utility of cultural replicators.

---

[16] In contrast, genetic replicators primarily depend upon vertical transmission of information, albeit with notable exceptions for bacterial plasmid exchange—which increases the rate at which they can adapt to novel conditions—as well as retro-virus mediated trans-individual sequence insertion (Tetz, 2005).

Nonetheless, because the success of cultural replicators is conditioned upon their ability to persist within nervous systems, overall cultural evolution is constrained by the utility dynamics of individuals. More specifically, to the extent that agents are "rational," they will tend to invest resources in learning cultural information and transmitting it to the degree that they believe that these activities are relevant to their values. Since the emergent preferences of individual nervous systems will tend to be canalized in ways that increase genetic fitness, biological evolution is able to influence cultural evolution indirectly via biasing 'action' selection.[17]

Thus, in addition to the processes of cultural group selection previously described, biology shapes and constrains culture in further ways. Because of these constraints, just as genetic replicators were able to increase their fitness (on average) by relying upon neural replicators, genetic fitness was enhanced (on average) to an even greater extent by reliance upon cultural replicators. Although the power of cultural optimization has been primarily discussed in terms of human-like cultural evolution, even sub-symbolic and non-imitative proto-cultures greatly enhance the power of learning processes. To the extent that dynamics of interaction are capable of persisting over time, optimized complexity can accrue, and thus individual animals are likely to benefit from social learning to the degree that it can occur (Whiten, McGuigan, Marshall-Pescini, & Hopper, 2009).

In terms of the analysis of "engineered" phenomena previously discussed, scientists encounter a challenging "reverse inference problem" (Poldrack, 2006, 2008, 2011). That is, since multiple causal pathways could lead to a given outcome, it is frequently unclear which selective pressures were operative in producing an adaptation. For similar reasons, it is extremely difficult to infer the extent and ways in which a specific aspect of behavior has been shaped by evolution on the level of genes, cultural factors, or gene-culture interactions.

However, by using details of mechanistic implementation as converging lines of evidence, it is possible to make increasingly precise and reliable inferences. In the next section, we will discuss some of the ways that vertebrate nervous systems have been shaped in order to ensure that preferences on the levels of individual organisms tend to coincide with genetic fitness (on average). When viewed in terms of multi-level evolutionary optimization, one can begin to understand the mystery of how the limited complexity of the genome and epigenome are capable of ensuring adaptive functioning of a network with billions of heterogeneous neurons, each with thousands of connections (or more).

## THE NEURAL SYSTEMS BASIS OF MEDO

---

[17] From an MEDO perspective, cognition and behavior necessarily have blurry boundaries, and actions are selected based on a variety of processes that may have little to do with specific motor sequences.

## NEUROANATOMY AND PHYLOGENY

*"All models are wrong, but some are useful."*

–George E. P. Box (1919 - present)

MacLean's "triune brain," provides a useful but over-simplified model that analyzes the vertebrate nervous system in terms of a prehistoric reptilian brain, to which an ancient mammalian brain was added, which then provided the foundation for a more recently evolved neo-mammalian brain (MacLean, 1983; Ploog, 2003). The justification for this proposal derived from observations that nervous system homology—for both structure and function—is more extensive for phylogenetically older structures, and decreases in a graded fashion for structures that evolved more recently. It also corresponded with the intuition that rather than abandoning successful physiological adaptations, natural selection would be more likely to modify existing 'designs' by either making incremental alterations to existing structures, or by adding new structures with new functions.

Additionally, the symptomatology of certain neuropsychiatric conditions—where developmentally acquired aspects of behavioral and emotional control were lost—seemed to support the 19th century maxim of "ontogeny recapitulates phylogeny" (Northcutt, 1990). However, this model is also limited in that newly developed structures interact with older structures and thereby change their properties, sometimes to an extent that phylogenetically older structures have diverged as they became optimized to increase adaptive functioning under these new configurations (Butler & Hodos, 2005). Nonetheless, older levels of organization provide necessary scaffolding for newer levels to work, both in terms of evolution, development, and ongoing functioning (E. N. Brown, Lydic, & Schiff, 2010; Långsjö et al., 2012; Merker, 2007; Panksepp, 1998, 2011).

## ACTION SELECTION AND THE BRAIN

*"Nothing in biology makes sense except in the light of evolution."*

–Theodosius Dobzhansky (1900-1975)

Although all organisms face the challenge of creating coherent functioning from heterogeneous elements, the spinal column and brainstem were the first major selection systems to evolve in vertebrates (Humphries et al., 2007). All of the external and internal sensory systems of the body converge on the reticular nuclei of the brainstem, which in turn project to numerous structures and can influence movement via cranial nerves and spinal motor pools. This convergence of inputs and divergence of outputs provides a centralized location in which various patterns can compete and cooperate in determining the behavior of the organism.

With respect to selecting between autonomic and neuroendocrine states, the hypothalamus is another structure with evolutionarily ancient origins (Hoebel, 1979; Jackson, 1981). The hypothalamus can be subdivided on the basis of anatomical features and functional properties, such as specific nuclei that mediate phylogenetically relevant behaviors (Pfaff & Sakuma, 1979). Further, through the pituitary gland, convergent inputs to the hypothalamus allow for coherent, large-scale physiological changes in response to the overall state of the organism (Toni, Malaguti, Benfenati, & Martini, 2004).

In addition to their other functions, brainstem and hypothalamic nuclei provide fundamental sources of neuroendocrine and autonomic control, which are basic types of 'action' selection. These structures are most clearly linked to biologically essential processes such as spinal-column reflexes for harm avoidance, brainstem nuclei modulating heart rate and respiration, or hypothalamic nuclei regulating "drives" for eating, drinking, or reproduction. In these ways, evolutionarily ancient substrates exist for enabling basic emotions, which can be thought of as coordinated changes in the body-brain state of the organism that prepare it to respond to a particular class of situations (A. D. Craig, 2003; A. D. B. Craig, 2004, 2009, 2011; A. Damasio, 2000, 2003; Panksepp, 1998). At their highest levels of abstraction, emotions are prototypes—or exemplar collections—of configurations of an organism's self-representations, corresponding to classes of situations encountered in the process of engaging in goal-oriented behavior.

These adaptations provide a means by which genetic natural selection was capable of influencing the primary sources of utility that shape evolution within individual organisms (Hall et al., 2000; Le Magnen, 1998; Pfaus et al., 2012). For example, if a pattern of activity results in successful feeding behaviors, projections from hypothalamic nuclei can stimulate neuromodulatory nuclei of the brainstem—where a single neuron can impact hundreds of thousands to millions of other neurons—which can then strengthen the connections between cell assemblies that contributed to that behavior (Butler & Hodos, 2005). To the extent that specific behaviors are "hard wired" into the brain, these "fixed action patterns" primarily consist of stereotyped muscular and neuroendocrine responses (e.g. the arching of the back involved in the lordosis mating reflex) (Eaton, Lee, & Foreman, 2001; Lin et al., 2011; Pfaff & Sakuma, 1979; Sokolowski & Corbin, 2012).

However, neural networks are inherently plastic, and thus the precise impacts of these primary reinforcers and action pattern generators will change through experience in even the simplest vertebrate nervous systems (Bolhuis, 1999). A particularly important factor influencing these shaping experiences will be the particular affordances and constraints associated with different body plans in different environments (Clark, 2008; Gibson, 1977; Hirose, 2002; Windridge & Kittler, 2010). In specifying particular aspects of embodiment, genes have yet another way of shaping utility dynamics within the organism. Thus, just as the functional significance of genes are determined by their ability to influence the multi-level dynamics that create phenotypes, the functional significance of neural adaptations is determined by their ability to influence dynamics on multiple levels both internal and external to the body. Also, similarly to how epigenetic mechanisms allow experience to shape gene expression—and thus increase phenotypic plasticity—neural systems are also capable of being shaped by the dynamics they influence.

Edelman's characterization of the processes biasing neural evolution as "values" fits well with a EDT analysis of neuropsychological phenomena (Wyckoff, 1987). To the extent that genetically specified mechanisms are able to influence selection within nervous systems, they contribute to the processes that determine the fitness of neural replicators. The stronger the selective pressures, the more likely it is that different aspects of neurophysiological functioning will have been shaped by these sources of utility. However, multiple needs must be satisfied in order for an organism to survive and reproduce, which themselves constitute values that are not always compatible (J. A. Miller, 2007). As previously described, since these different values form part of a common system, there is a sense in which each value competes and cooperates with every other value in shaping dynamics of multi-objective optimization.

In the Darwinian struggle to shape utility dynamics, mechanisms that are more clearly "innate" have at least three important advantages over mechanisms that are more dependent upon learning. First, to the extent that these mechanisms are important for ensuring survival and reproduction (e.g. swallowing), specific details of neurophysiology may allow these adaptations to be particularly robust. Second, if these biologically essential mechanisms require frequent and precise modulation (e.g. amount of drinking), then there will be many opportunities for these selective pressures to shape dynamics. Third, if a selective pressure is capable of operating early in development (e.g. desire for water), it has a "first mover advantage" in shaping overall dynamics, and thus it may be able to maintain a dominant position relative to other values. However, even for behaviors as fundamental to survival as drinking, experience is required for newborn rats to learn the association between approaching water and obtaining hydration when thirsty (Hall et al., 2000).

Over the course of evolution, additional adaptations allowed for increasingly complicated utility dynamics to take place within nervous systems. The nuclei of the basal ganglia (BG) are crucial structures for enabling selection between potential motor sequences, which are capable of being modified on the basis of experience (Gurney et al., 2001; Houk et al., 2007). BG neurons form segregated "loops" of mutually inhibitory connections with external structures, thus helping to ensure that only robust ensembles of neural patterns are capable of inhibiting their tonic inhibition. This disinhibitory configuration encourages "winner-take-all" dynamics by preventing weaker patterns from interfering with stronger patterns (Mao & Massaquoi, 2007; Rutishauser, Douglas, & Slotine, 2010), thus helping to ensure the functional coherence and stability of the collective neural activity underlying perception-action cycles.

In humans, BG structures include the caudate, putamen, globus pallidus, substantia nigra, and subthalamic nuclei. In the caudate and putamen, collectively known as the striatum, the more dorsal division receives dopaminergic inputs from the substantia nigra and is more closely associated with specific motoric processes, while the more ventral division receives dopaminergic inputs from the ventral tegmentum and is associated with motivation more generally. Activations in the ventral striatum—which in humans, also refers to the dopaminergic projections to the nucleus accumbens and subcallosal cortex—are often interpreted as indicating "reward", along with other constructs such as prediction error, attention, or salience (K. S. Smith, Berridge, & Aldridge, 2011). Thus, reinforcement learning is more appropriately understood as a systems property of the brain that can involve many neural processes having little to do with specific hedonic experiences.

The relative ease with which patterns are capable of disinhibiting themselves is highly dependent on dopamine levels within the basal ganglia (Gerfen & Surmeier, 2011; Haruno

& Kawato, 2006; Paladini, Celada, & Tepper, 1999). Multiple areas—including homeostatic regulatory nuclei—converge on dopaminergic neuromodulatory nuclei of the midbrain, and thus there is a sense in which fluctuating dopamine levels represent a kind of "common neural currency" that integrates different forms of utility (Leknes & Tracey, 2008; Watson, 2008). However, since the conditions under which this currency increases or decreases depends on predicted outcomes (Bray & O'Doherty, 2007; Glimcher, 2011; Valentin & O'Doherty, 2009), these utility dynamics change as organisms learn through experience. Also, while these mechanisms play a key role in determining which behaviors are likely to be rewarded (Berridge, 2006), reward learning can occur in mice that have been genetically engineered to lack dopamine (Robinson, Sandstrom, Denenberg, & Palmiter, 2005). Thus, it is important to keep in mind that there is no single criterion of selection within the brain, and there may even be "competing" neural currencies.

These neural systems arose over 560 million years ago with important aspects of functioning being conserved in all vertebrates (Butler & Hodos, 2005; Stephenson-Jones, Ericsson, Robertson, & Grillner, 2012). However, these structures also became modified as expanded nervous system complexity provided them with new functional significances.

## CORTICAL CONTROL SYSTEMS

*"There is nothing more practical than a good theory."*

   –Kurt Lewin (1890-1947)

The evolution of cortex allowed organisms to remember specific sequences of events from the past, and on the basis of these memories, predict specific sequences that might occur in the future. In Fuster's (2006) model of cortex, memory and knowledge are represented in distributed and overlapping networks of interacting neurons that are "heterarchically" organized into functional cycles within perception-action hierarchies (Joaquin M Fuster, 2007; Joaquín M Fuster, 2006, 2009; Joaquín M Fuster & Bressler, 2012). He refers to these auto-associated representational units as "cognits," and considers them to be the elemental basis of cortical memory. Over the course of development, increasingly sophisticated representations evolve as more complex patterns are abstracted from simpler patterns, which allow new kinds of patterns to be discovered. From this perspective, nested dynamics of causal relations in the world are mirrored by nested dynamics of neuronal representations, which exist in varying degrees of abstraction and complexity with respect to perception-action sequences.

Intuitively, hierarchical organization suggests that this gradient of increasing abstraction is reflected by locations of representations relative to primary sensory and motor cortices. That is, concrete perception-action sequences will be located closer to where sensations first input, and where specific motor commands first output. More abstract patterns (e.g. high-level goals) will be located closer to multi-modal association cortices and further from primary cortices. However, these abstract patterns will be connected to representations on multiple levels of the hierarchy on the basis of auto-associative links formed through past and ongoing experience.

Converging support for this theory is beginning to accumulate from the fields of computational neuroscience and machine learning. More specifically, it has been proposed that cortex may implement a common algorithm for hierarchical pattern abstraction, which affords complex associations that would be difficult to discover by statistical learning within non-hierarchically organized networks (Friston, 2008; Friston & Kiebel, 2009; George & Hawkins, 2009; Hawkins & Blakeslee, 2004; Kiebel, Daunizeau, & Friston, 2008; Schofield et al., 2009). In addition to its theoretical parsimony, this hypothesis is supported by studies of cortical micro-circuitry. After discovering the existence of cortical columns in 1978, Mountcastle suggested these structures potentially enabled a common computational process to take place throughout cortex (Mountcastle, 1978). More recently, machine learning systems are being developed in which hierarchically organized cortical columns constitute nodes within probabilistic networks for advanced pattern recognition (Hawkins, 2011).

Cortex is anatomically distinguished by groups of approximately 80-100 neurons with common developmental origin, arranged into columns that are positioned perpendicularly to a cortical sheet with six semi-distinct layers (Krieger, Kuner, & Sakmann, 2007; Mountcastle, 1997).[18] Except for a few differences in visual and motor cortices, these "minicolumns" show remarkable anatomical homogeneity across brain areas, as well as across different mammalian species. Notably, cytoarchitectonic characterization suggests that morphological variation is initially lacking, and diversity is created by experience (Buxhoeveden & Casanova, 2002; E. G. Jones & Rakic, 2010). Minicolumns further self-organize into groups of approximately 60-100 units to form macrocolumns (H. Markram, 2006). These macrocolumns, or "cortical modules" consist of multiple minicolumns bound together via short-range connections with similar, although still heterogeneous, response properties.[19] The total number of cortical modules for an average human brain is difficult to precisely determine, with estimates ranging from one to two million columns (Johansson & Lansner, 2007). Although its specific functional importance has been the subject of debate (Horton & Adams, 2005)—partially due to inconsistent use of terminology—columnar organization is a fundamental feature of cortex.

Theoretically, cortex could efficiently select for utility maximizing sequences by minimizing the "free energy" of the underlying processes (Friston, 2010; Hawkins, 2011; Kozma, Puljic,

---

[18] Although the "limbic" structures of the amygdala and hippocampus, as well as the "para-limbic" structures of the cingulate and insula are commonly referred to as "subcortical," technically, they are all different kinds of cortex (Mesulam & Mufson, 1982). The amygdala and hippocampus have a three-layered structure, the surrounding areas of the amygdala and the parahippocampal cortex have a four-layer structure, the cingulate and insula have a five-layer structure, and neo-cortex has a six-layer structure. These different layers reflect the order that these structures mature over the course of development, as well as the order that they developed over the course of evolution, and are called archicortex, peri-allocortex, and iso-cortex, respectively. The significance of these different numbers of layers is currently unknown.

[19] Columnar modules have no relation or resemblance to the special-purpose computational modules discussed within evolutionary psychology.

Balister, Bollobas, & Freeman, 2004). If a minicolumn's inputs are predicted in advance via stimulation of specific inhibitory interneurons within the column, then only those neurons without their respective inhibitory interneurons activated will increase their firing rates. However, if a sufficient number of non-predicted inputs occur, and a percolation threshold is surpassed, the entire column may become active, resulting in a cascade of activity-predictions in functionally connected columns. Depending on the degree of functional connectivity with inhibitory interneurons of midbrain neuromodulatory nuclei (Watabe-Uchida, Zhu, Ogawa, Vamanrao, & Uchida, 2012), any dynamic that causes overall activity to be reduced should result in decreased inhibition of the production of these neuromodulators. This net disinhibition would enhance the most robustly active patterns, strengthen the connections underlying these patterns (i.e. reinforcement), and thus increase the efficiency of the dynamics contributing to successful prediction (i.e. minimized error signals).

Although this activity-minimizing algorithm could potentially result in stasis, regulatory nuclei of the hypothalamus and midbrain would stimulate these inhibitory interneurons to the degree that action is needed to restore homeostatic balances. Thus an organism could not remain permanently inactive, as physiological signals such as hunger would result in stimulation of these regulatory nuclei, whose activity can be thought of as signifying the distance from homeostatic set points, or as signifiers of biologically specified predictions for which deviations result in error signals. Over time, cortical dynamics resulting in the minimization of error signals from these regulatory nuclei will become distributed across the cortical heterarchy as habitual predictions. The impact of these habitual predictions on overall functioning would constitute the evolving utility function of the organism.

It is difficult to overstate the importance of the body in providing an initial set of constraints and values for the evolving nervous system. Indeed, organisms may be such effective learners because they begin with a sense of their own embodiment as a kind of prototypical object from which they can partially generalize, and they pay attention to this object because it is directly connected to the mechanisms of reinforcement (Leknes & Tracey, 2008). This account of cognitive development can potentially explain the knowledge that infants possess (Carey & Gelman, 1991), as well as how they use it to partially generalize to other classes of phenomena, even sometimes incorrectly (Lakoff & Johnson, 1999).

Importantly, the functional significance of a particular cortical column or neuron emerges as a function of the unique experiences of an individual organism. Thus, it is unlikely that natural selection was capable of shaping behavior through genetically specified cortical representations, or responses to those representations. A priori, the non-linearity of development suggests that genetic selection would be unable to produce such adaptations, even with strong selective pressures. If a mechanism relies on pre-specifying details of biology that are unpredictable in principle (Elbert et al., 1994; Korn & Faure, 2003; Rabinovich & Abarbanel, 1998; Wolfram, 2002), then it is also un-evolvable. Functional predetermination is possible on the basis of cellular characteristics (e.g. excitatory vs. inhibitory, degree of dendritic/axonal branching), locations of neurons with specific cortical lamina, as well as relative positioning with respect to specific sensory inputs and outputs to various effector systems. However, since the formation of specific connections is based on information that could not be known in advance, according to Edelman, all neurons are "gypsies" in the developing brain (Edelman, 1992). Exceptions seem to be rare, such as brainstem reticular neurons involved in the "escape reflex" (Eaton et al., 2001).

Thus in maximizing fitness, genes necessarily relied upon additional optimization processes, which they could only influence indirectly, utilizing mechanisms for innate value setting to bias neuronal evolution. For example, connections between homeostatic regulatory nuclei and brainstem neurons controlling neuromodulator levels help to ensure that an organisms actions will be reinforced or punished based on their ability to contribute to survival.

Although the existence of reliably predictable associations between cortical areas and cognitive functions may seem to challenge these claims, these regularities can also be explained in terms of experience shaping systems that self-organize through evolutionary processes. Indeed, inter-individual variability and developmental evidence suggest these associations are not genetically specified in any direct sense (Pataraia et al., 2004; Peelen, Glaser, Vuilleumier, & Eliez, 2009).

However, to the extent that a hierarchical structure increases the predictability of cortical organization, it is probable that some degree of genetic accommodation has occurred. For example, it is not a coincidence that the area associated with language comprehension is located close to where auditory information first enters cortex. It is also not a coincidence that this area is just enough steps removed where intermediate cortex can divide time varying auditory sequences into phonemes, morphemes, syllables, and finally words. Nor is it a coincidence that this area is located close to "association cortex" where converging high-level patterns from multiple sensory modalities allow the meaning of symbols to be grounded. Again, since perception and action are intimately linked, it is not coincidental that this language comprehension area is located near speech production areas, which are themselves unsurprisingly located near areas of the motor-strip that control the tongue and mouth, as well as areas of the frontal lobes that support high-level goal representations. Indeed, none of this is surprising when the cortex is viewed as a heterarchical memory system for perception and action, grounded in the sensors and effectors of the body.

To the extent that cortex could be expected to reliably self-organize in a way where these particular functions have these specific relative locations, genes would be likely to facilitate these processes by influencing any detail of physiology that would allow for evolving neural systems to more efficiently reach these fitness maxima. This sort of canalization can be realized through multiple neurodevelopmental pathways. Thus, neuropsychological phenomena can be genetically influenced to a significant degree, even if it is difficult to identify specific genotypic configurations that significantly contribute to the frequency of particular phenotypes. Mechanistic flexibility could potentially extend the range of phenomena with "innate" bases, but the inherent non-linearity of self-organizing systems limits predictability. This developmental uncertainty reduces the precision with which natural selection could specify which canals are likely to emerge within the extended epigenetic landscapes that shape evolving neural systems.

Although the most obvious change in cortex to take place over the course of evolution is that cerebral volumes have greatly increased, this expansion has been most pronounced in association cortices, particularly in humans (Preuss, 2011). Expanding the cortical heterarchy allowed for increasingly complicated and abstracted patterns (Premack, 1983), which as previously described, afforded enhanced social cognition and eventually cultural evolution. These additional dynamics greatly expanded the non-linearity of human development, which necessarily forced genetic replicators to increasingly rely upon general-purpose learning mechanisms for ensuring adaptive behavior. Further, to the extent that these dynamics are constrained by innately specified values, the information they provide will be increasingly reliable for enhancing genetic fitness (on average). Thus, even though "innate" factors contribute to determining behavior in all organisms, humans are unique in the extent to which they depend upon learning.

## REFERENCES

Abbas, A. E. (2010). General Decompositions of Multiattribute Utility Functions with Partial Utility Independence. *Journal of MultiCriteria Decision Analysis*, *59*(November 2009), 37–59. doi:10.1002/mcda

Aldrich, H., Hodgson, G., Hull, D., Knudsen, T., Mokyr, J., & Vanberg, V. (2008). In defence of generalized Darwinism. *Journal of Evolutionary Economics*, *18*(5), 577–596. doi:10.1007/s00191-008-0110-z

Allport, D. A. (1985). Distributed memory, modular subsystems and dysphasia (pp. 32–60).

Anfora, G., Rigosi, E., Frasnelli, E., Ruga, V., Trona, F., & Vallortigara, G. (2011). Lateralization in the invertebrate brain: left-right asymmetry of olfaction in bumble bee, Bombus terrestris. *PloS One*, *6*(4), e18903. doi:10.1371/journal.pone.0018903

Atran, S. (2001). The trouble with memes: Inference versus imitation in cultural creation. *Human Nature*, *12*, 351–381.

Attwell, D., & Laughlin, S. B. (2001). An energy budget for signaling in the grey matter of the brain. *Journal of Cerebral Blood Flow and Metabolism: Official Journal of the International Society of Cerebral Blood Flow and Metabolism*, *21*(10), 1133–1145. doi:10.1097/00004647-200110000-00001

Ayala, F. J. (2007). Colloquium Papers: Darwin's greatest discovery: Design without designer. *Proceedings of the National Academy of Sciences*, *104*(suppl_1), 8567–8573. doi:10.1073/pnas.0701072104

Badyaev, A. V. (2009). Evolutionary significance of phenotypic accommodation in novel environments: an empirical test of the Baldwin effect. *Philosophical Transactions of the Royal Society of London. Series B, Biological Sciences*, *364*(1520), 1125–1141. doi:10.1098/rstb.2008.0285

Bak, P, & Paczuski, M. (1995). Complexity, contingency, and criticality. *Proceedings of the National Academy of Sciences of the United States of America*, *92*(15), 6689–6696.

Bak, Per, Tang, C., & Wiesenfeld, K. (1987). Self-organized criticality: An explanation of the 1/f noise. *Physical Review Letters*, *59*(4), 381–384. doi:10.1103/PhysRevLett.59.381

Baldwin, J. M. (1896). A New Factor in Evolution. *The American Naturalist*, *30*(354), 441–451. doi:10.1086/276408

Barbieri, M. (2008). Biosemiotics: a new understanding of life. *Die Naturwissenschaften*, *95*(7), 577–599. doi:10.1007/s00114-008-0368-x

Barsalou, L. W. (2008). Grounded cognition. *Annual Review of Psychology*, *59*, 617–645. doi:10.1146/annurev.psych.59.103006.093639

Bawazer, L. A., Izumi, M., Kolodin, D., Neilson, J. R., Schwenzer, B., & Morse, D. E. (2012). Evolutionary Selection of Enzymatically Synthesized Semiconductors from Biomimetic Mineralization Vesicles. *Proceedings of the National Academy of Sciences*. doi:10.1073/pnas.1116958109

Begon, M., Harper, J. L., & Townsend, C. R. (1999). *Ecology: Individuals, Populations and Communities* (3rd ed.). Wiley-Blackwell.

Bell, A. (2010). Why cultural and genetic group selection are unequal partners in the evolution of human behavior. *Communicative & Integrative Biology*, *3*(2), 159–161.

Bell, D. E. (1979). Multiattribute Utility Functions: Decompositions Using Interpolation. *Management Science*, *25*(8), 744–753. doi:10.1287/mnsc.25.8.744

Berger, J. O. (1985). *Statistical decision theory and Bayesian analysis*. Springer. Retrieved from http://books.google.com/books?hl=en&lr=&id=oY_x7dE15_AC&oi=fnd&pg=PA1&dq=%22bayesian+decision+theory%22&ots=wAJ4ubp0T6&sig=HjZdCHP61qcyI7OqwPCAuhD-Rvc

Berridge, K. C. (2006). The debate over dopamine's role in reward: the case for incentive salience. *Psychopharmacology*, *191*(3), 391–431. doi:10.1007/s00213-006-0578-x

Bolhuis, J. J. (1999). The development of animal behavior: from Lorenz to neural nets. *Die Naturwissenschaften*, *86*(3), 101–111.

Both, S., Spiering, M., Laan, E., Belcome, S., van den Heuvel, B., & Everaerd, W. (2008). Unconscious classical conditioning of sexual arousal: evidence for the conditioning of female genital arousal to subliminally presented sexual stimuli. *The Journal of Sexual Medicine*, *5*(1), 100–109. doi:10.1111/j.1743-6109.2007.00643.x

Boyd, & Richerson, P. J. (2000). Meme theory oversimplifies how culture changes. *Scientific American*, *283*(4), 70–71.

Bray, S., & O'Doherty, J. (2007). Neural coding of reward-prediction error signals during classical conditioning with attractive faces. *Journal of Neurophysiology*, *97*(4), 3036–3045. doi:10.1152/jn.01211.2006

Bridgeman, B., & Tseng, P. (2011). Embodied cognition and the perception-action link. *Physics of Life Reviews*, *8*(1), 73–85. doi:10.1016/j.plrev.2011.01.002

Brown, E. N., Lydic, R., & Schiff, N. D. (2010). General anesthesia, sleep, and coma. *The New England Journal of Medicine*, *363*(27), 2638–2650. doi:10.1056/NEJMra0808281

Butler, A. B., & Hodos, W. (2005). *Comparative Vertebrate Neuroanatomy: Evolution and Adaptation* (2nd ed.). Wiley-Liss.

Buxhoeveden, D. P., & Casanova, M. F. (2002). The minicolumn hypothesis in neuroscience. *Brain: A Journal of Neurology*, *125*(Pt 5), 935–951.

Call, J., & Tomasello, M. (2008). Does the chimpanzee have a theory of mind? 30 years later. *Trends in Cognitive Sciences*, *12*(5), 187–192. doi:10.1016/j.tics.2008.02.010

Campbell, J. (2009). Bayesian Methods and Universal Darwinism. *AIP Conference Proceedings*, *1193*(1), 40–47. doi:doi:10.1063/1.3275642

Caporale, N., & Dan, Y. (2008). Spike timing-dependent plasticity: a Hebbian learning rule. *Annual Review of Neuroscience*, *31*, 25–46. doi:10.1146/annurev.neuro.31.060407.125639

Carey, S., & Gelman, R. (1991). *The Epigenesis of Mind: Essays on Biology and Cognition*. Psychology Press.

Clark, A. (2008). *Supersizing the Mind: Embodiment, Action, and Cognitive Extension* (1st ed.). Oxford University Press, USA.

Coello, C. A. C., Dehuri, S., & Ghosh, S. (2009). *Swarm Intelligence For Multi-Objective Problems in Data Mining*. Springer.

Craig, A. D. (2003). Interoception: the sense of the physiological condition of the body. *Current Opinion in Neurobiology*, *13*(4), 500–505.

Craig, A. D. B. (2004). Human feelings: why are some more aware than others? *Trends in Cognitive Sciences*, *8*(6), 239–241. doi:10.1016/j.tics.2004.04.004

Craig, A. D. B. (2009). How do you feel--now? The anterior insula and human awareness. *Nature Reviews. Neuroscience*, *10*(1), 59–70. doi:10.1038/nrn2555

Craig, A. D. B. (2011). Significance of the insula for the evolution of human awareness of feelings from the body. *Annals of the New York Academy of Sciences*, *1225*, 72–82. doi:10.1111/j.1749-6632.2011.05990.x

Crispo, E. (2007). The Baldwin effect and genetic assimilation: revisiting two mechanisms of evolutionary change mediated by phenotypic plasticity. *Evolution; International Journal of Organic Evolution*, *61*(11), 2469–2479. doi:10.1111/j.1558-5646.2007.00203.x

Damasio, A. (2000). *The Feeling of What Happens: Body and Emotion in the Making of Consciousness* (1st ed.). Mariner Books.

Damasio, A. (2003). *Looking for Spinoza: Joy, Sorrow, and the Feeling Brain* (1st ed.). Houghton Mifflin Harcourt.

Davies, A. P., Watson, R. A., Mills, R., Buckley, C. L., & Noble, J. (2011). "If you can't be with the one you love, love the one you're with": how individual habituation of agent interactions improves global utility. *Artificial Life*, *17*(3), 167–181. doi:10.1162/artl_a_00030

Dawkins, R. (1983). Universal darwinism. Retrieved from http://www.citeulike.org/group/2050/article/1379902

Dawkins, Richard. (1976). *The Selfish Gene* (2nd ed.). Oxford University Press, USA.

Dawkins, Richard. (1996a). *River Out Of Eden: A Darwinian View Of Life*. Basic Books.

Dawkins, Richard. (1996b). *The Blind Watchmaker: Why the Evidence of Evolution Reveals a Universe without Design*. W. W. Norton & Company.

Dawkins, Richard. (1997). *Climbing Mount Improbable*. W. W. Norton & Company.

Dawkins, Richard. (1999). *The Extended Phenotype: The Long Reach of the Gene* (Revised.). Oxford University Press, USA.

Day, R. L., Laland, K. N., & Odling-Smee, F. J. (2003). Rethinking adaptation: the niche-construction perspective. *Perspectives in Biology and Medicine*, *46*(1), 80–95.

Dennett, D. (1992). *Consciousness Explained* (1st ed.). Back Bay Books.

Dennett, D. (1996). *DARWIN'S DANGEROUS IDEA: EVOLUTION AND THE MEANINGS OF LIFE*. Simon & Schuster.

Dennett, D. (2003). *Freedom Evolves* (illustrated edition.). Viking Adult.

Dennett, D. (2009). Colloquium Papers: Darwin's "strange inversion of reasoning." *Proceedings of the National Academy of Sciences*, *106*(Supplement_1), 10061–10065. doi:10.1073/pnas.0904433106

Eaton, R. C., Lee, R. K. K., & Foreman, M. B. (2001). The Mauthner cell and other identified neurons of the brainstem escape network of fish. *Progress in Neurobiology*, *63*(4), 467–485. doi:10.1016/S0301-0082(00)00047-2

Eckert, A. J., & Dyer, R. J. (2012). Defining the landscape of adaptive genetic diversity. *Molecular Ecology*, *21*(12), 2836–2838. doi:10.1111/j.1365-294X.2012.05615.x

Edelman, G. (1992). *Bright Air, Brilliant Fire: On The Matter Of The Mind*. Basic Books.

Edelman, G. (1993). Neural Darwinism: Selection and reentrant signaling in higher brain function. *Neuron*, *10*(2), 115–125. doi:10.1016/0896-6273(93)90304-A

Edelman, G., & Mountcastle, V. B. (1978). *The Mindful Brain: Cortical Organization and the Group-Selective Theory of Higher Brain Function* (1st ed.). MIT Press.

Elbert, T., Ray, W. J., Kowalik, Z. J., Skinner, J. E., Graf, K. E., & Birbaumer, N. (1994). Chaos and physiology: deterministic chaos in excitable cell assemblies. *Physiological Reviews*, *74*(1), 1–47.

Eldredge, & Gould, S. J. (1972). Punctuated equilibria: an alternative to phyletic gradualism, *82*(5), 82–115. doi:10.1037/h0022328

Elliott, T., Maddison, A. C., & Shadbolt, N. R. (2001). Competitive anatomical and physiological plasticity: a neurotrophic bridge. *Biological Cybernetics*, *84*(1), 13–22.

Ernande, B., & Dieckmann, U. (2004). The evolution of phenotypic plasticity in spatially structured environments: implications of intraspecific competition, plasticity costs and environmental characteristics. *Journal of Evolutionary Biology*, *17*(3), 613–628. doi:10.1111/j.1420-9101.2004.00691.x

Faratin, P., Klein, M., Sayama, H., & Bar-Yam, Y. (2002). Simple Negotiating Agents in Complex Games: Emergent Equilibria and Dominance of Strategies. In J.-J. Meyer & M. Tambe (Eds.), *Intelligent Agents VIII* (Vol. 2333, pp. 367–376). Springer Berlin / Heidelberg. Retrieved from http://www.springerlink.com/content/6xv96rutcfjgc9q1/abstract/

Fay, J. C., Wyckoff, G. J., & Wu, C.-I. (2001). Positive and Negative Selection on the Human Genome. *Genetics*, *158*(3), 1227–1234.

Foley, R. (1995). The adaptive legacy of human evolution: A search for the environment of evolutionary adaptedness. *Evolutionary Anthropology: Issues, News, and Reviews*, *4*(6), 194–203. doi:10.1002/evan.1360040603

Fracchia, J., & Lewontin, R. C. (2005). The Price of Metaphor. *History and Theory, History and Theory, 44, 44*(1, 1), 14, 14–29, 29. doi:10.1111/j.1468-2303.2005.00305.x, 10.1111/j.1468-2303.2005.00305.x

Friston, K. (2008). Hierarchical models in the brain. *PLoS Computational Biology*, *4*(11), e1000211. doi:10.1371/journal.pcbi.1000211

Friston, K. (2010). The free-energy principle: a unified brain theory? *Nature Reviews. Neuroscience*, *11*(2), 127–138. doi:10.1038/nrn2787

Friston, K., & Kiebel, S. (2009). Predictive coding under the free-energy principle. *Philosophical Transactions of the Royal Society of London. Series B, Biological Sciences*, *364*(1521), 1211–1221. doi:10.1098/rstb.2008.0300

Fuster, Joaquin M. (2007). Jackson and the frontal executive hierarchy. *International Journal of Psychophysiology: Official Journal of the International Organization of Psychophysiology*, *64*(1), 106–107. doi:10.1016/j.ijpsycho.2006.07.014

Fuster, Joaquín M. (2006). The cognit: a network model of cortical representation. *International Journal of Psychophysiology: Official Journal of the International Organization of Psychophysiology*, *60*(2), 125–132. doi:10.1016/j.ijpsycho.2005.12.015

Fuster, Joaquín M. (2009). Cortex and memory: emergence of a new paradigm. *Journal of Cognitive Neuroscience*, *21*(11), 2047–2072. doi:10.1162/jocn.2009.21280

Fuster, Joaquín M, & Bressler, S. L. (2012). Cognit activation: a mechanism enabling temporal integration in working memory. *Trends in Cognitive Sciences*. doi:10.1016/j.tics.2012.03.005

Galef, B. G. (2009). What can function tell us about mechanism? *Trends in Ecology & Evolution*, *24*(7), 357–358. doi:10.1016/j.tree.2009.02.002

Galuske, R. A., Kim, D. S., Castrén, E., & Singer, W. (2000). Differential effects of neurotrophins on ocular dominance plasticity in developing and adult cat visual cortex. *The European Journal of Neuroscience*, *12*(9), 3315–3330.

Galuske, R. A., Kim, D. S., Castren, E., Thoenen, H., & Singer, W. (1996). Brain-derived neurotrophic factor reversed experience-dependent synaptic modifications in kitten visual cortex. *The European Journal of Neuroscience*, *8*(7), 1554–1559.

Gause, G. F. (1934). *The Struggle for Existence*. Courier Dover Publications.

Geisler, W. S., & Diehl, R. L. (2002). Bayesian natural selection and the evolution of perceptual systems. *Philosophical Transactions of the Royal Society of London. Series B, Biological Sciences*, *357*(1420), 419–448. doi:10.1098/rstb.2001.1055

George, D., & Hawkins, J. (2009). Towards a mathematical theory of cortical micro-circuits. *PLoS Computational Biology*, *5*(10), e1000532. doi:10.1371/journal.pcbi.1000532

Gerfen, C. R., & Surmeier, D. J. (2011). Modulation of Striatal Projection Systems by Dopamine. *Annual Review of Neuroscience*, *34*(1), 441–466. doi:10.1146/annurev-neuro-061010-113641

Gibson, J. J. (1977). The theory of affordances (Vol. Perceiving, pp. 67–82).

Gilad, Y., Oshlack, A., Smyth, G. K., Speed, T. P., & White, K. P. (2006). Expression profiling in primates reveals a rapid evolution of human transcription factors. *Nature*, *440*(7081), 242–245. doi:10.1038/nature04559

Gilbert, D. (2006). *Stumbling on Happiness* (1st ed.). Knopf.

Glimcher, P. W. (2011). Understanding dopamine and reinforcement learning: the dopamine reward prediction error hypothesis. *Proceedings of the National Academy*

*of Sciences of the United States of America*, *108 Suppl 3*, 15647–15654. doi:10.1073/pnas.1014269108

Gogtay, N., Giedd, J. N., Lusk, L., Hayashi, K. M., Greenstein, D., Vaituzis, A. C., … Thompson, P. M. (2004). Dynamic Mapping of Human Cortical Development During Childhood Through Early Adulthood. *Proceedings of the National Academy of Sciences of the United States of America*, *101*(21), 8174–8179. doi:10.1073/pnas.0402680101

Gould, S. J. (1990). *Wonderful Life: The Burgess Shale and the Nature of History*. W. W. Norton & Company.

Gould, S. J. (1992). *The Panda's Thumb: More Reflections in Natural History*. W. W. Norton & Company.

Gould, S. J., & Lewontin, R. J. (1979). The spandrels of San Marco and the Panglossian paradigm: a critique of the adaptationist programme. *Proceedings of the Royal Society of London. Series B, Containing Papers of a Biological Character. Royal Society (Great Britain)*, *205*(1161), 581–598.

Gurney, K., Prescott, T. J., & Redgrave, P. (2001). A computational model of action selection in the basal ganglia. I. A new functional anatomy. *Biological Cybernetics*, *84*(6), 401–410.

Gw, T., W, F., D, F., & M, B. (1995). Multi-attribute preference functions. Health Utilities Index. *PharmacoEconomics*, *7*(6), 503.

Hall, W. G., Arnold, H. M., & Myers, K. P. (2000). The acquisition of an appetite. *Psychological Science*, *11*(2), 101–105.

Haruno, M., & Kawato, M. (2006). Heterarchical reinforcement-learning model for integration of multiple cortico-striatal loops: fMRI examination in stimulus-action-reward association learning. *Neural Networks: The Official Journal of the International Neural Network Society*, *19*(8), 1242–1254. doi:10.1016/j.neunet.2006.06.007

Haslberger, A., Varga, F., & Karlic, H. (2006). Recursive causality in evolution: a model for epigenetic mechanisms in cancer development. *Medical Hypotheses*, *67*(6), 1448–1454. doi:10.1016/j.mehy.2006.05.047

Hawkins. (2011). Hierarchical Temporal Memory: including HTM Cortical Learning Algorithms. *Whitepaper Numenta Inc*. Retrieved from http://www.numenta.com/htm-overview/education/HTM_CorticalLearningAlgorithms.pdf

Hawkins, J., & Blakeslee, S. (2004). *On Intelligence* (Adapted.). Times Books.

Hayek, F. A. (1952). *The Sensory Order: An Inquiry into the Foundations of Theoretical Psychology*. University Of Chicago Press.

Hebb, D. O. (1949). *The Organization of Behavior: A Neuropsychological Theory* (New edition.). Psychology Press.

Hennigan, A., O'Callaghan, R. M., & Kelly, A. M. (2007). Neurotrophins and their receptors: roles in plasticity, neurodegeneration and neuroprotection. *Biochemical Society Transactions*, *35*(Pt 2), 424–427. doi:10.1042/BST0350424

Henrich, J. (2004). Cultural group selection, coevolutionary processes and large-scale cooperation. *Journal of Economic Behavior & Organization*, *53*(1), 3–35. doi:10.1016/S0167-2681(03)00094-5

Heylighen, F. (1992). *Principles of Systems and Cybernetics: an evolutionary perspective*.

Heylighen, F. (1999). *The Growth of Structural and Functional Complexity during Evolution*.

Heylighen, F., & Joslyn, C. (2003). Cybernetics and Second-Order Cybernetics. In *Encyclopedia of Physical Science and Technology (Third Edition)* (pp. 155–169). New York: Academic Press. Retrieved from http://www.sciencedirect.com/science/article/pii/B0122274105001617

Hirose. (2002). An ecological approach to embodiment and cognition. *Cognitive Systems Research*, *3*(3), 289–299. doi:10.1016/S1389-0417(02)00044-X

Hoebel, B. G. (1979). Hypothalamic self-stimulation and stimulation escape in relation to feeding and mating. *Federation Proceedings*, *38*(11), 2454–2461.

Hoffmann, H., Janssen, E., & Turner, S. L. (2004). Classical conditioning of sexual arousal in women and men: effects of varying awareness and biological relevance of the conditioned stimulus. *Archives of Sexual Behavior*, *33*(1), 43–53. doi:10.1023/B:ASEB.0000007461.59019.d3

Hofstadter, D. R. (1979). *Godel, Escher, Bach: An Eternal Golden Braid* (Vol. 14). Basic Books.

Holland, J. H. (1992). Adaptation in Natural and Artificial Systems: An Introductory Analysis with Applications to Biology, Control, and Artificial Intelligence. Retrieved from http://mitpress.mit.edu/catalog/item/default.asp?ttype=2&tid=8929

Honig, L. S., & Rosenberg, R. N. (2000). Apoptosis and neurologic disease. *The American Journal of Medicine*, *108*(4), 317–330.

Horton, J. C., & Adams, D. L. (2005). The cortical column: a structure without a function. *Philosophical Transactions of the Royal Society B: Biological Sciences*, *360*(1456), 837–862. doi:10.1098/rstb.2005.1623

Houk, J. C., Bastianen, C., Fansler, D., Fishbach, A., Fraser, D., Reber, P. J., … Simo, L. S. (2007). Action selection and refinement in subcortical loops through basal ganglia and cerebellum. *Philosophical Transactions of the Royal Society of London. Series B, Biological Sciences*, *362*(1485), 1573–1583. doi:10.1098/rstb.2007.2063

House, P. K., Vyas, A., & Sapolsky, R. (2011). Predator Cat Odors Activate Sexual Arousal Pathways in Brains of Toxoplasma gondii Infected Rats. *PLoS ONE*, *6*(8), e23277. doi:10.1371/journal.pone.0023277

Howarth, C., Gleeson, P., & Attwell, D. (2012). Updated energy budgets for neural computation in the neocortex and cerebellum. *Journal of Cerebral Blood Flow and Metabolism: Official Journal of the International Society of Cerebral Blood Flow and Metabolism*. doi:10.1038/jcbfm.2012.35

Huang, S. (2012). The molecular and mathematical basis of Waddington's epigenetic landscape: a framework for post-Darwinian biology? *BioEssays: News and Reviews in Molecular, Cellular and Developmental Biology*, *34*(2), 149–157. doi:10.1002/bies.201100031

Humphries, M. ., Gurney, K., & Prescott, T. . (2007). Is there a brainstem substrate for action selection? *Philosophical Transactions of the Royal Society B: Biological Sciences*, *362*(1485), 1627 –1639. doi:10.1098/rstb.2007.2057

Jablonka, E., & Lamb, M. J. (2007). Précis of Evolution in Four Dimensions. *The Behavioral and Brain Sciences*, *30*(4), 353–365; discusssion 365–389. doi:10.1017/S0140525X07002221

Jackson, I. M. (1981). Evolutionary significance of the phylogenetic distribution of the mammalian hypothalamic releasing hormones. *Federation Proceedings*, *40*(11), 2545–2552.

James, W. (1890). *Psychology, briefer course*. Franklin Library.

JämsÄ, T. (2001). Jakob von Uexkülls theory of sign and meaning from a philosophical, semiotic, and linguistic point of view. *Semiotica*, *2001*(134), 481–551. doi:10.1515/semi.2001.042

Johansson, C., & Lansner, A. (2007). Towards cortex sized artificial neural systems. *Neural Networks*, *20*(1), 48–61. doi:10.1016/j.neunet.2006.05.029

Johnston, T. D., & Gottlieb, G. (1990). Neophenogenesis: a developmental theory of phenotypic evolution. *Journal of Theoretical Biology*, *147*(4), 471–495.

Jones, B. D. (1999). Bounded Rationality. *Annual Review of Political Science*, *2*(1), 297–321. doi:10.1146/annurev.polisci.2.1.297

Jones, E. G., & Rakic, P. (2010). Radial Columns in Cortical Architecture: It Is the Composition That Counts. *Cerebral Cortex*, *20*(10), 2261–2264. doi:10.1093/cercor/bhq127

Jones, P. A., & Takai, D. (2001). The Role of DNA Methylation in Mammalian Epigenetics. *Science*, *293*(5532), 1068–1070. doi:10.1126/science.1063852

Kahneman, D., & Tversky, A. (1979). Prospect theory: An analysis of decision under risk. *Econometrica: Journal of the Econometric Society*, 263–291.

Kahneman, Daniel. (2003). A perspective on judgment and choice: mapping bounded rationality. *American Psychologist*, *58*(9), 697–720.

Kahneman, Daniel. (2011). *Thinking, Fast and Slow* (1st ed.). Farrar, Straus and Giroux.

Kahneman, Daniel, & Tversky, A. (2007). Prospect Theory: An Analysis of Decision under Risk Daniel Kahneman; Amos Tversky. *Econometrica*, *47*(2), 263–292.

Kaminski, J., Call, J., & Tomasello, M. (2008). Chimpanzees know what others know, but not what they believe. *Cognition*, *109*(2), 224–234. doi:10.1016/j.cognition.2008.08.010

Keeney, R. L., & Raiffa, H. (1993). *Decisions with Multiple Objectives: Preferences and Value Tradeoffs*. Cambridge University Press.

Kiebel, S. J., Daunizeau, J., & Friston, K. J. (2008). A hierarchy of time-scales and the brain. *PLoS Computational Biology*, *4*(11), e1000209. doi:10.1371/journal.pcbi.1000209

King, A. C., & Bjorklund, D. F. (2010). Evolutionary developmental psychology. *Psicothema*, *22*(1), 22–27.

Kohl, J. V. (2012). Human pheromones and food odors: epigenetic influences on the socioaffective nature of evolved behaviors. *Socioaffective Neuroscience & Psychology*, *2*(0). doi:10.3402/snp.v2i0.17338

Korn, H., & Faure, P. (2003). Is there chaos in the brain? II. Experimental evidence and related models. *Comptes Rendus Biologies*, *326*(9), 787–840.

Kozma, R., Puljic, M., Balister, P., Bollobas, B., & Freeman, W. (2004). Neuropercolation: A Random Cellular Automata Approach to Spatio-temporal Neurodynamics. In P. Sloot, B. Chopard, & A. Hoekstra (Eds.), *Cellular Automata* (Vol. 3305, pp. 435–443). Springer Berlin / Heidelberg. Retrieved from http://www.springerlink.com/content/jq3d3uj89p9ql7cf/abstract/

Krieger, P., Kuner, T., & Sakmann, B. (2007). Synaptic connections between layer 5B pyramidal neurons in mouse somatosensory cortex are independent of apical dendrite bundling. *The Journal of Neuroscience: The Official Journal of the Society for Neuroscience*, *27*(43), 11473–11482. doi:10.1523/JNEUROSCI.1182-07.2007

Lakoff, G., & Johnson, M. (1999). *Philosophy in the Flesh : The Embodied Mind and Its Challenge to Western Thought*. Basic Books.

Laland, K. N., Odling-Smee, F. J., & Feldman, M. W. (1999). Evolutionary consequences of niche construction and their implications for ecology. *Proceedings of the National Academy of Sciences of the United States of America*, *96*(18), 10242–10247.

Lane, N. (2005). *Power, Sex, Suicide: Mitochondria and the Meaning of Life* (1st ed.). Oxford University Press, USA.

Långsjö, J. W., Alkire, M. T., Kaskinoro, K., Hayama, H., Maksimow, A., Kaisti, K. K., … Scheinin, H. (2012). Returning from Oblivion: Imaging the Neural Core of Consciousness. *The Journal of Neuroscience*, *32*(14), 4935–4943. doi:10.1523/JNEUROSCI.4962-11.2012

Lansing, J. S., & Fox, K. M. (2011). Niche construction on Bali: the gods of the countryside. *Philosophical Transactions of the Royal Society of London. Series B, Biological Sciences*, *366*(1566), 927–934. doi:10.1098/rstb.2010.0308

Larsson, T. A., Olsson, F., Sundstrom, G., Lundin, L.-G., Brenner, S., Venkatesh, B., & Larhammar, D. (2008). Early vertebrate chromosome duplications and the evolution of the neuropeptide Y receptor gene regions. *BMC Evolutionary Biology*, *8*, 184. doi:10.1186/1471-2148-8-184

Le Magnen, J. (1998). Synthetic approach to the neurobiology of behaviour. *Appetite*, *31*(1), 1–8. doi:10.1006/appe.1998.0176

Lefebvre, L. (2012). Primate encephalization. *Progress in Brain Research*, *195*, 393–412. doi:10.1016/B978-0-444-53860-4.00019-2

Leknes, S., & Tracey, I. (2008). A common neurobiology for pain and pleasure. *Nature Reviews. Neuroscience*, *9*(4), 314–320. doi:10.1038/nrn2333

Lessells, C. K. M. (2008). Neuroendocrine control of life histories: what do we need to know to understand the evolution of phenotypic plasticity? *Philosophical Transactions of the Royal Society of London. Series B, Biological Sciences*, *363*(1497), 1589–1598. doi:10.1098/rstb.2007.0008

Lewis, K. E., Chen, W., & Schmidt, L. C. (Eds.). (2006). *Decision Making in Engineering Design*. Three Park Avenue New York, NY 10016-5990: ASME. Retrieved from http://asmedl.org/ebooks/asme/asme_press/802469/802469_ch12

Lin, D., Boyle, M. P., Dollar, P., Lee, H., Perona, P., Lein, E. S., & Anderson, D. J. (2011). Functional identification of an aggression locus in the mouse hypothalamus. *Nature*, *470*(7333), 221–226. doi:10.1038/nature09736

Lovecchio, E., Allegrini, P., Geneston, E., West, B. J., & Grigolini, P. (2012). From self-organized to extended criticality. *Frontiers in Physiology*, *3*, 98. doi:10.3389/fphys.2012.00098

Lynch, M. (2010). Evolution of the mutation rate. *Trends in Genetics: TIG*, *26*(8), 345–352. doi:10.1016/j.tig.2010.05.003

Lynch, M. (2011). The lower bound to the evolution of mutation rates. *Genome Biology and Evolution*, *3*, 1107–1118. doi:10.1093/gbe/evr066

MacLean, P. D. (1983). BRAIN ROOTS OF THE WILL-TO-POWER. *Zygon®*, *18*(4), 359–374. doi:10.1111/j.1467-9744.1983.tb00522.x

Mao, Z.-H., & Massaquoi, S. G. (2007). Dynamics of winner-take-all competition in recurrent neural networks with lateral inhibition. *IEEE Transactions on Neural Networks / a Publication of the IEEE Neural Networks Council*, *18*(1), 55–69. doi:10.1109/TNN.2006.883724

Markram, H. (2006). The blue brain project. *Nature Reviews. Neuroscience*, *7*(2), 153–160. doi:10.1038/nrn1848

Markram, K., & Markram, H. (2010). The intense world theory - a unifying theory of the neurobiology of autism. *Frontiers in Human Neuroscience*, *4*, 224. doi:10.3389/fnhum.2010.00224

Marshall, A. (1925). *Principles of economics: an introductory volume*. Macmillan and Company, limited.

Maynard Smith. (1982). *Evolution and Theory of Games* (Vol. 64). Cambridge University Press.

Maynard Smith, J., & Szathmáry, E. (1995). *The Major Transitions in Evolution* (Vol. 19). Oxford University Press.

McDowell, J. J. (2010). Behavioral and neural Darwinism: selectionist function and mechanism in adaptive behavior dynamics. *Behavioural Processes*, *84*(1), 358–365. doi:10.1016/j.beproc.2009.11.011

McNally, L., Brown, S. P., & Jackson, A. L. (2012). Cooperation and the Evolution of Intelligence. *Proceedings of the Royal Society B: Biological Sciences*. doi:10.1098/rspb.2012.0206

McNally, R. J. (1987). Preparedness and phobias: a review. *Psychological Bulletin*, *101*(2), 283–303.

McNally, R. J. (1995). Preparedness, phobias, and the Panglossian paradigm. *Behavioral and Brain Sciences*, *18*(2), 303–304.

Merker, B. (2007). Consciousness without a cerebral cortex: a challenge for neuroscience and medicine. *The Behavioral and Brain Sciences*, *30*(1), 63–81; discussion 81–134. doi:10.1017/S0140525X07000891

Mesulam, M. M., & Mufson, E. J. (1982). Insula of the old world monkey. I. Architectonics in the insulo-orbito-temporal component of the paralimbic brain. *The Journal of Comparative Neurology*, *212*(1), 1–22. doi:10.1002/cne.902120102

Miller, J. A. (2007). Repeated evolution of male sacrifice behavior in spiders correlated with genital mutilation. *Evolution; International Journal of Organic Evolution*, *61*(6), 1301–1315. doi:10.1111/j.1558-5646.2007.00115.x

Mills, R., & Watson, R. A. (2005). Genetic Assimilation and Canalisation in the Baldwin Effect. *Advances in Artificial Life*.

Mineka, S., & Ohman, A. (2002). Phobias and preparedness: the selective, automatic, and encapsulated nature of fear. *Biological Psychiatry*, *52*(10), 927–937.

Misevic, D., Kouyos, R. D., & Bonhoeffer, S. (2009). Predicting the Evolution of Sex on Complex Fitness Landscapes. *PLoS Comput Biol*, *5*(9), e1000510. doi:10.1371/journal.pcbi.1000510

Morris, J. S., Ohman, A., & Dolan, R. J. (1998). Conscious and unconscious emotional learning in the human amygdala. *Nature*, *393*(6684), 467–470. doi:10.1038/30976

Mountcastle, V. B. (1978). An organizing principle for cerebral function. (pp. 7–50).

Mountcastle, V. B. (1997). The columnar organization of the neocortex. *Brain: A Journal of Neurology*, *120 ( Pt 4)*, 701–722.

Murray, E. K., Hien, A., de Vries, G. J., & Forger, N. G. (2009). Epigenetic control of sexual differentiation of the bed nucleus of the stria terminalis. *Endocrinology*, *150*(9), 4241–4247. doi:10.1210/en.2009-0458

Nelson, R. (2007). Universal Darwinism and evolutionary social science. *Biology and Philosophy*, *22*(1), 73–94. doi:10.1007/s10539-005-9005-7

Northcutt, R. G. (1990). Ontogeny and phylogeny: a re-evaluation of conceptual relationships and some applications. *Brain behavior and evolution*, *36*(2-3), 116–140.

Nowak, M. A. (2006). Five rules for the evolution of cooperation. *Science (New York, N.Y.)*, *314*(5805), 1560–1563. doi:10.1126/science.1133755

Nowak, M. A., & Sigmund, K. (2004). Evolutionary dynamics of biological games. *Science (New York, N.Y.)*, *303*(5659), 793–799. doi:10.1126/science.1093411

Odling-Smee, F. J., Laland, K. N., & Feldman, M. W. (2003). *Niche Construction: The Neglected Process in Evolution*. Princeton University Press.

Ohman, A., & Mineka, S. (2001). Fears, phobias, and preparedness: toward an evolved module of fear and fear learning. *Psychological Review*, *108*(3), 483–522.

Paladini, C. A., Celada, P., & Tepper, J. M. (1999). Striatal, pallidal, and pars reticulata evoked inhibition of nigrostriatal dopaminergic neurons is mediated by GABA(A) receptors in vivo. *Neuroscience*, *89*(3), 799–812.

Panksepp, J. (1998). *Affective Neuroscience: The Foundations of Human and Animal Emotions* (illustrated edition.). Oxford University Press, USA.

Panksepp, J. (2011). Cross-species affective neuroscience decoding of the primal affective experiences of humans and related animals. *PLoS One*, *6*(9), e21236. doi:10.1371/journal.pone.0021236

Patalano, S., Hore, T. A., Reik, W., & Sumner, S. (2012). Shifting behaviour: epigenetic reprogramming in eusocial insects. *Current Opinion in Cell Biology*, *24*(3), 367–373. doi:10.1016/j.ceb.2012.02.005

Pataraia, E., Simos, P. G., Castillo, E. M., Billingsley-Marshall, R. L., McGregor, A. L., Breier, J. I., … Papanicolaou, A. C. (2004). Reorganization of Language-Specific Cortex in Patients with Lesions or Mesial Temporal Epilepsy. *Neurology*, *63*(10), 1825–1832. doi:10.1212/01.WNL.0000144180.85779.9A

Peelen, M. V., Glaser, B., Vuilleumier, P., & Eliez, S. (2009). Differential development of selectivity for faces and bodies in the fusiform gyrus. *Developmental Science*, *12*(6), F16–25. doi:10.1111/j.1467-7687.2009.00916.x

Pfaff, D. W., & Sakuma, Y. (1979). Facilitation of the lordosis reflex of female rats from the ventromedial nucleus of the hypothalamus. *The Journal of Physiology*, *288*, 189–202.

Pfaus, J. G., Kippin, T. E., Coria-Avila, G. A., Gelez, H., Afonso, V. M., Ismail, N., & Parada, M. (2012). Who, What, Where, When (and Maybe Even Why)? How the Experience of Sexual Reward Connects Sexual Desire, Preference, and Performance. *Archives of Sexual Behavior*, *41*(1), 31–62. doi:10.1007/s10508-012-9935-5

Pianka, E. R. (1970). On r-and K-selection. *The American Naturalist*, *104*(940), 592–597.

Pigliucci, M. (2009). An extended synthesis for evolutionary biology. *Annals of the New York Academy of Sciences*, *1168*, 218–228. doi:10.1111/j.1749-6632.2009.04578.x

Pine, A., Seymour, B., Roiser, J. P., Bossaerts, P., Friston, K. J., Curran, H. V., & Dolan, R. J. (2009). Encoding of marginal utility across time in the human brain. *The Journal of Neuroscience: The Official Journal of the Society for Neuroscience*, *29*(30), 9575–9581. doi:10.1523/JNEUROSCI.1126-09.2009

Ploog, D. W. (2003). The place of the Triune Brain in psychiatry. *Physiology & Behavior*, *79*(3), 487–493.

Poldrack, R. A. (2006). Can cognitive processes be inferred from neuroimaging data? *Trends in Cognitive Sciences*, *10*(2), 59–63. doi:10.1016/j.tics.2005.12.004

Poldrack, R. A. (2008). The role of fMRI in cognitive neuroscience: where do we stand? *Current Opinion in Neurobiology*, *18*(2), 223–227. doi:10.1016/j.conb.2008.07.006

Poldrack, R. A. (2011). The future of fMRI in cognitive neuroscience. *NeuroImage*. doi:10.1016/j.neuroimage.2011.08.007

Premack, D. (1983). The Codes of Man and Beasts. *Behavioral and Brain Sciences*, *6*(01), 125–136. doi:10.1017/S0140525X00015077

Preuss, T. M. (2011). The human brain: rewired and running hot. *Annals of the New York Academy of Sciences*, *1225 Suppl 1*, E182–191. doi:10.1111/j.1749-6632.2011.06001.x

Price, T. D., Qvarnström, A., & Irwin, D. E. (2003). The role of phenotypic plasticity in driving genetic evolution. *Proceedings. Biological Sciences / The Royal Society*, *270*(1523), 1433–1440. doi:10.1098/rspb.2003.2372

Rabinovich, M. I., & Abarbanel, H. D. (1998). The role of chaos in neural systems. *Neuroscience*, *87*(1), 5–14.

Reader, S. M., Hager, Y., & Laland, K. N. (2011). The evolution of primate general and cultural intelligence. *Philosophical Transactions of the Royal Society of London. Series B, Biological Sciences*, *366*(1567), 1017–1027. doi:10.1098/rstb.2010.0342

Rendell, L., Fogarty, L., & Laland, K. N. (2011). Runaway cultural niche construction. *Philosophical Transactions of the Royal Society of London. Series B, Biological Sciences*, *366*(1566), 823–835. doi:10.1098/rstb.2010.0256

Reznick, D., Bryant, M. J., & Bashey, F. (2002). r- AND K-SELECTION REVISITED: THE ROLE OF POPULATION REGULATION IN LIFE-HISTORY EVOLUTION. *Ecology*, *83*(6), 1509–1520. doi:10.1890/0012-9658(2002)083[1509:RAKSRT]2.0.CO;2

Richter, H. (2010). Evolutionary Optimization and Dynamic Fitness Landscapes. In I. Zelinka, S. Celikovsky, H. Richter, & G. Chen (Eds.), (Vol. 267, pp. 409–446). Springer Berlin / Heidelberg. Retrieved from http://www.springerlink.com/content/m222l00356u72425/abstract/

Ridley, M. (1993). *The red queen: Sex and the evolution of human nature*. Harper Perennial. Retrieved from http://books.google.com/books?hl=en&lr=&id=7AypMwk_TD0C&oi=fnd&pg=PA245&dq=ideas+have+sex+ridley&ots=RsQTSp9ook&sig=NQtGW4vlfq_JeDzm6X-_KWXegos

Robinson, S., Sandstrom, S. M., Denenberg, V. H., & Palmiter, R. D. (2005). Distinguishing whether dopamine regulates liking, wanting, and/or learning about rewards. *Behavioral Neuroscience*, *119*(1), 5–15. doi:10.1037/0735-7044.119.1.5

Rudrauf, D., Lutz, A., Cosmelli, D., Lachaux, J.-P., & Le Van Quyen, M. (2003). From autopoiesis to neurophenomenology: Francisco Varela's exploration of the biophysics of being. *Biological Research*, *36*(1), 27–65.

Rutishauser, U., Douglas, R. J., & Slotine, J.-J. (2010). Collective Stability of Networks of Winner-Take-All Circuits. *Neural Computation*. doi:10.1162/NECO_a_00091

Samuelson. (1938). A Note on the Pure Theory of Consumer's Behaviour. *Economica*, *51*(17). Retrieved from http://dx.doi.org/10.2307/2548836

Sawaragi, Y., Nakayama, H., & Tanino, T. (1985). *Theory of multiobjective optimization*. Academic Press.

Schofield, T. M., Iverson, P., Kiebel, S. J., Stephan, K. E., Kilner, J. M., Friston, K. J., … Leff, A. P. (2009). Changing meaning causes coupling changes within higher levels of the cortical hierarchy. *Proceedings of the National Academy of Sciences of the United States of America*, *106*(28), 11765–11770. doi:10.1073/pnas.0811402106

Schott, G. D. (2011). Freud's Project and Its Diagram: Anticipating the Hebbian Synapse. *Journal of Neurology, Neurosurgery & Psychiatry*, *82*(2), 122–125. doi:10.1136/jnnp.2010.220400

Scott, J. (2010). Rational Choice Theory (Vol. 1, pp. 1–11).

Segal, N. L., & MacDonald, K. B. (1998). Behavioral genetics and evolutionary psychology: unified perspective on personality research. *Human Biology*, *70*(2), 159–184.

Seligman, M. E. P. (1971). Phobias and preparedness. *Behavior Therapy*, *2*(3), 307–320. doi:10.1016/S0005-7894(71)80064-3

Seth, A. K., & Baars, B. J. (2005). Neural Darwinism and consciousness. *Consciousness and Cognition*, *14*(1), 140–168. doi:10.1016/j.concog.2004.08.008

Shatz, C. J. (1996). Emergence of order in visual system development. *Proceedings of the National Academy of Sciences of the United States of America*, *93*(2), 602–608.

Shay, J. (2010). From an unlicensed philosopher: reflections on brain, mind, society, culture--each other's environments with equal "ontologic standing." *Annals of the New York Academy of Sciences*, *1208*, 32–37. doi:10.1111/j.1749-6632.2010.05797.x

Shorter, J. (2010). Emergence and natural selection of drug-resistant prions. *Molecular bioSystems*, *6*(7), 1115–1130. doi:10.1039/c004550k

Simon, H. (1955). A Behavioral Model of Rational Choice. *Quarterly Journal of Economics*, *69*(1), 99–118. doi:10.2307/1884852

Simon, H., Egidi, M., Viale, R., & Marris, R. L. (2008). *Economics, Bounded Rationality and the Cognitive Revolution*. Edward Elgar Publishing.

Sisk, C. L., & Foster, D. L. (2004). The neural basis of puberty and adolescence. *Nature Neuroscience*, *7*(10), 1040–1047. doi:10.1038/nn1326

Skinner, B. F. (1938). The behavior of organisms: An experimental analysis. Retrieved from http://psycnet.apa.org/psycinfo/1939-00056-000

Slack, J. M. W. (2002). Conrad Hal Waddington: the last Renaissance biologist? *Nature Reviews. Genetics*, *3*(11), 889–895. doi:10.1038/nrg933

Smit, S. K., & Eiben, A. E. (2009). Comparing parameter tuning methods for evolutionary algorithms. In *IEEE Congress on Evolutionary Computation, 2009. CEC '09* (pp. 399 – 406). Presented at the IEEE Congress on Evolutionary Computation, 2009. CEC '09. doi:10.1109/CEC.2009.4982974

Smith, J. M. (1976). Evolution and the theory of games. *American Scientist*, *64*(1), 41–45.

Smith, K. S., Berridge, K. C., & Aldridge, J. W. (2011). Disentangling pleasure from incentive salience and learning signals in brain reward circuitry. *Proceedings of the National Academy of Sciences of the United States of America*, *108*(27), E255–264. doi:10.1073/pnas.1101920108

Sokolowski, K., & Corbin, J. G. (2012). Wired for behaviors: from development to function of innate limbic system circuitry. *Frontiers in Molecular Neuroscience*, *5*. doi:10.3389/fnmol.2012.00055

Song, S., Miller, K. D., & Abbott, L. F. (2000). Competitive Hebbian learning through spike-timing-dependent synaptic plasticity. *Nature Neuroscience*, *3*(9), 919–926. doi:10.1038/78829

Stella, M., & Kleisner, K. (2010). Uexküllian Umwelt as science and as ideology: the light and the dark side of a concept. *Theory in Biosciences = Theorie in Den Biowissenschaften*, *129*(1), 39–51. doi:10.1007/s12064-010-0081-0

Stephenson-Jones, M., Ericsson, J., Robertson, B., & Grillner, S. (2012). Evolution of the basal ganglia; Dual output pathways conserved throughout vertebrate phylogeny. *The Journal of Comparative Neurology*. doi:10.1002/cne.23087

Struhsaker, T. T., Cooney, D. O., & Siex, K. S. (1997). Charcoal Consumption by Zanzibar Red Colobus Monkeys: Its Function and Its Ecological and Demographic Consequences. *International Journal of Primatology*, *18*(1), 61–72. doi:10.1023/A:1026341207045

Suzuki, R., & Arita, T. (2007). The dynamic changes in roles of learning through the Baldwin effect. *Artificial Life*, *13*(1), 31–43. doi:10.1162/artl.2007.13.1.31

Takeuchi, N., Hogeweg, P., & Koonin, E. V. (2011). On the origin of DNA genomes: evolution of the division of labor between template and catalyst in model replicator systems. *PLoS Computational Biology*, *7*(3), e1002024. doi:10.1371/journal.pcbi.1002024

Takeuchi, N., Salazar, L., Poole, A. M., & Hogeweg, P. (2008). The evolution of strand preference in simulated RNA replicators with strand displacement: implications for the origin of transcription. *Biology Direct*, *3*, 33. doi:10.1186/1745-6150-3-33

Tetz, V. V. (2005). The pangenome concept: a unifying view of genetic information. *Medical Science Monitor: International Medical Journal of Experimental and Clinical Research*, *11*(7), HY24–29.

Thompson, E. (2006). *Neurophenomenology and Francisco Varela*. Harvard University Press.

Thompson, N. S. (1998). Reintroducing "Reintroducing group selection to the human behavioral sciences"to BBS readers. *Behavioral and Brain Sciences*, *21*(02), 304–305. doi:null

Tinbergen, N. (1963). On aims and methods of Ethology. *Zeitschrift für Tierpsychologie*, *20*(4), 410–433. doi:10.1111/j.1439-0310.1963.tb01161.x

Tomasello, M. (1999). THE HUMAN ADAPTATION FOR CULTURE. *Annual Review of Anthropology*, *28*(1), 509–529. doi:10.1146/annurev.anthro.28.1.509

Toni, R., Malaguti, A., Benfenati, F., & Martini, L. (2004). The human hypothalamus: a morpho-functional perspective. *Journal of Endocrinological Investigation*, *27*(6 Suppl), 73–94.

Tooby, J., & Cosmides, L. (1990). On the universality of human nature and the uniqueness of the individual: the role of genetics and adaptation. *Journal of Personality*, *58*(1), 17–67.

Traulsen, A., & Nowak, M. A. (2006). Evolution of cooperation by multilevel selection. *Proceedings of the National Academy of Sciences of the United States of America*, *103*(29), 10952–10955. doi:10.1073/pnas.0602530103

Trepel, C., Fox, C. R., & Poldrack, R. A. (2005). Prospect theory on the brain? Toward a cognitive neuroscience of decision under risk. *Brain Research*, *23*(1), 34–50.

Ullén, F. (2009). Is activity regulation of late myelination a plastic mechanism in the human nervous system? *Neuron Glia Biology*, *5*(1-2), 29–34. doi:10.1017/S1740925X09990330

Valentin, V. V., & O'Doherty, J. P. (2009). Overlapping prediction errors in dorsal striatum during instrumental learning with juice and money reward in the human brain. *Journal of Neurophysiology*, *102*(6), 3384–3391. doi:10.1152/jn.91195.2008

Van Rossum, M. C., Bi, G. Q., & Turrigiano, G. G. (2000). Stable Hebbian learning from spike timing-dependent plasticity. *The Journal of Neuroscience: The Official Journal of the Society for Neuroscience*, *20*(23), 8812–8821.

Von Neumann. (1966). *Theory of Self-Reproducing Automata*. University of Illinois Press.

Von Neumann, & Morgenstern, O. (1944). *Theory of Games and Economic Behavior* (Vol. 2). Princeton University Press.

Von Uexküll, J. (1957). A stroll through the worlds of animals and men (Vol. 4, pp. 5–80).

Waddington, C. H. (1953a). Genetic assimilation of an acquired character. *Evolution*, *7*(2), 118–126. doi:10.2307/2405747

Waddington, C. H. (1953b). The'Baldwin Effect,'Genetic Assimilation'and'Homeostasis'. *Evolution*, *7*(4), 386–387.

Wagner, G. P., & Laubichler, M. D. (2004). Rupert Riedl and the re-synthesis of evolutionary and developmental biology: body plans and evolvability. *Journal of Experimental Zoology. Part B, Molecular and Developmental Evolution*, *302*(1), 92–102. doi:10.1002/jez.b.20005

Watabe-Uchida, M., Zhu, L., Ogawa, S. K., Vamanrao, A., & Uchida, N. (2012). Whole-Brain Mapping of Direct Inputs to Midbrain Dopamine Neurons. *Neuron*, *74*(5), 858–873. doi:10.1016/j.neuron.2012.03.017

Waters, D. P. (2011). Von Neumann's Theory of Self-Reproducing Automata: A Useful Framework for Biosemiotics? *Biosemiotics*, *onlin 25 J*. doi:10.1007/s12304-011-9127-z

Watson, K. K. (2008). Evolution, Risk, and Neural Representation. *Annals of the New York Academy of Sciences*, *1128*(1), 8–12. doi:10.1196/annals.1399.002

Whiten, A., McGuigan, N., Marshall-Pescini, S., & Hopper, L. M. (2009). Emulation, imitation, over-imitation and the scope of culture for child and chimpanzee. *Philosophical Transactions of the Royal Society of London. Series B, Biological Sciences*, *364*(1528), 2417–2428. doi:10.1098/rstb.2009.0069

Williams, G. C. (1966). *Adaptation and Natural Selection: A Critique of Some Current Evolutionary Thought*. Princeton University Press.

Wilson, D. S. (1994). Reintroducing group selection to the human behavioral sciences. *Behavioral and Brain Sciences*, *17*(04), f1–f8. doi:10.1017/S0140525X00036086

Windridge, D., & Kittler, J. (2010). Perception-action learning as an epistemologically-consistent model for self-updating cognitive representation. *Advances in Experimental Medicine and Biology*, *657*, 95–134. doi:10.1007/978-0-387-79100-5_6

Wolfram, S. (2002). *A New Kind of Science* (1st ed.). Wolfram Media.

Wright. (1932). The roles of mutation, inbreeding, crossbreeding and selection in evolution. *Proceedings of the Sixth International Congress of Genetics*, *1*(6), 356–366.

Wright, S. (1982). The shifting balance theory and macroevolution. *Annual Review of Genetics*, *16*, 1–19. doi:10.1146/annurev.ge.16.120182.000245

Wyckoff, G. J. (1987). *Neural Darwinism: The Theory Of Neuronal Group Selection* (First Edition.). Basic Books.

Ziemke, T., & Sharkey, N. E. (2001). A stroll through the worlds of robots and animals: Applying Jakob von Uexkülls theory of meaning to adaptive robots and artificial life. *Semiotica*, *2001*(134), 701–746. doi:10.1515/semi.2001.050